\documentstyle{elsart}    

\input psfig
\input epsf

\def\slc#1{\setbox0=\hbox{$#1$}  \dimen0=\wd0
   \setbox1=\hbox{/} \dimen1=\wd1
   \ifdim\dimen0>\dimen1   \rlap{\hbox to \dimen0{\hfil/\hfil}}  #1
   \else  \rlap{\hbox to \dimen1{\hfil$#1$\hfil}} /
   \fi}

\def\Nc{N_{\rm c}}

\def\xslide#1#2#3#4#5#6{\centerline{\psfig
{figure=#1,height=#2,bbllx=#3bp,bblly=#4bp,bburx=#5bp,bbury=#6bp,clip=}}}

\begin{document}

\thispagestyle{empty} ~ \vspace{-27mm}

\hfill{%
\parbox[t]{3.5in}{\begin{flushright}
                                                RUB-TPII-13/95\\
                             \end{flushright} }} \baselineskip=10pt

\vspace{0mm}

\begin{center}
{\LARGE {\bf Meson loops in the Nambu--Jona-Lasinio model}}

\vspace{6mm} {\large Emil N.~Nikolov$^a$, Wojciech Broniowski$^b$, Christo
V.~Christov$^a$, \\\vspace{3mm} Georges Ripka$^c$, and Klaus Goeke$^a$}

\vspace{9mm} {\sl $^a$ Institut f\"ur Theoretische Physik II,
Ruhr-Universit\"at Bochum,\\D-44780 Bochum, Germany}

\vspace{-2mm} {\sl $^b$ H. Niewodnicza\'nski Institute of Nuclear Physics,
PL-31342 Cracow, Poland}

\vspace{-2mm} {\sl $^c$ Service de Physique Th\'eorique, Centre d'Etudes de
Saclay, F-91191 Gif-sur-Yvette Cedex, France}
\end{center}

\vfill

\epsfxsize = 6 cm \centerline{\epsfbox{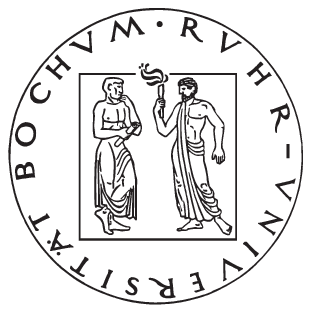}} \vspace{0mm}

\baselineskip=7pt

\begin{center}
{\Large Ruhr-Universit\"at Bochum \\Institut f\"ur Theoretische Physik II\\%
Teilchen- und Kernphysik }
\end{center}

\newpage
\begin{frontmatter}
\title{Meson loops in the Nambu-Jona-Lasinio model}
\author[Bochum]{Emil N.~Nikolov\thanksref{Sofia}},
\author[Krakow]{Wojciech Broniowski},
\author[Bochum]{Christo V.~Christov\thanksref{Sofia}},
\author[Saclay]{Georges Ripka},  and
\author[Bochum]{Klaus Goeke}
\address[Bochum]{Institut f\"ur Theoretische Physik II,
Ruhr-Universit\"at Bochum,\\
D-44780 Bochum, Germany}
\address[Krakow]{H. Niewodnicza\'nski Institute of Nuclear Physics,
PL-31342 Cracow, Poland}
\address[Saclay]{Service de Physique Th\'eorique de Saclay, F-91191
Gif-sur-Yvette Cedex, France}
\thanks[Sofia]{On leave of absence from
Institute for Nuclear Research and Nuclear Energy, 1784 Sofia, Bulgaria}
\begin{abstract}
The effects of meson loops in the vacuum sector of the
Nambu--Jona-Lasinio model are calculated. Using the 
effective action formalism
we take consistently all next-to-leading-order
 $1 \over \Nc$ terms into account.
This leads to a symmetry-conserving approach, in which all features
of spontaneously broken chiral symmetry, such as the Goldstone theorem, the
Gold\-ber\-ger--Treiman and 
the Gell-Mann--Oakes--Renner relations are preserved.
Contributions to $\langle \overline{q}q \rangle$
and $F_\pi$ are calculated, and are shown to be substantial, at the level of
 $\sim 30\%$, consistent with the  $1 \over \Nc$ expansion.
The leading nonanalytic terms in the chiral expansion of
 $\langle \overline{q}q \rangle$, $F_\pi$ and $m_\pi$ 
have the same form as the
one-loop results of chiral perturbation theory.
\end{abstract}
\end{frontmatter}

\footnotetext[2]{
E-mail:
\parbox[t]{4.5in}
{\baselineskip 3mm
emiln {\em or} christoc {\em or} goeke@hadron.tp2.ruhr-uni-bochum.de\\
broniows@solaris.ifj.edu.pl\\
ripka@amoco.saclay.cea.fr}}

\setcounter{page}{1}

\section{Introduction}

\label{sec:intro} Over the last few years the Nambu--Jona-Lasinio model~\cite
{NJL} has been used extensively to study numerous features of
strong-interaction physics at low energies (for recent reviews on mesons see~%
\cite{Weise,Vogl2,Bijnens:rev}, and on baryons see~\cite{Alkofer:rev,NJL:rev}%
). The four-fermion point interaction leads to dynamical chiral symmetry
breaking which plays a key role for the description of low-energy hadron
phenomena. An attractive feature of the model is that having a small number
of free parameters it allows for a unified description of mesons and
baryons, with both two and three flavors. The resulting hadron spectroscopy
has been quite successful: the basic mass relations, the electromagnetic
form factors and the electromagnetic polarizabilities are reproduced
reasonably well.

All the above applications have been performed in the {\em one-quark-loop}
approximation. The effective action, the meson and quark self-energies,
vertex functions, {\em etc.}, are all generated by a quark-loop. This
approximation is justified by the large-$N_{{\rm c}}$ limit, where meson
loops are $\frac{1}{N_{{\rm c}}}$-suppressed. However, meson loops cannot
always be neglected since in the real world $N_{{\rm c}}=3$. More
importantly, the pions are very light, and in many situations pionic
effects, although formally $\frac{1}{N_{{\rm c}}}$-suppressed, are enhanced
because of the low pion mass. For example, pionic thermal excitations
described by meson loops are dominating in hadronic systems at finite
temperature. Meson loops are also required for the proper description of
heavier mesons. For example, the $\sigma $ or the $\rho $ mesons can split
into two pions and this effect is again described by pionic loops~\cite
{Krewald:rho,Celenza93}.

There have been a number of attempts~\cite
{Cao92,Domitrovich93,Heid:loops,Tegen95} to include meson loops in the
Nambu--Jona-Lasinio model. These approaches are based on $\frac{1}{N_{{\rm c}%
}}$ expansion in terms of Feynman graphs. The symmetry properties of the
system, however, can be easily violated by an inappropriate choice of
diagrams. This is the case of Refs.~\cite{Cao92,Domitrovich93,Heid:loops}
where basic symmetry relations, such as {\em e.g.} the Goldstone theorem,
are violated. The reason for this is that the equation 
determining the expectation value of the $\sigma$ field and 
expressions for the meson
propagators are not consistent. We show that in order to guarantee
the validity of the Goldstone theorem on the basis of a consistent $\frac
1{N_{{\rm c}}}$ expansion it is crucial to determine both the 
expectation value of the $\sigma$ field
and the meson propagators consistently, {\em e.g.} using 
the effective action
including one-meson-loop contributions. 
Many-body methods which preserve the symmetry properties of the theory 
are called {\em symmetry-conserving approximations}~\cite
{Kadanoff,Georges}. The approach of Ref.~\cite{Tegen95} is up to our
knowledge the only one which preserves the Goldstone theorem at the one
meson-loop level. This is achieved by a judicious choice of Feynman diagrams.

In this paper we extend consistently the Nambu--Jona-Lasinio model to
next-to-leading order in $\frac{1}{N_{{\rm c}}}$ which includes the
meson-loop contributions. We use the effective action formalism which leads
in a natural way to a symmetry conserving approximation. Our approach
identifies the Feynman diagrams which need to be included to maintain the
symmetry properties, at any level of approximation. In our calculation the
pion remains massless in the chiral limit and the basic relations following
from Ward-Takahashi identities, such as the Gell-Mann--Oaks--Renner relation
and the Goldberger-Treiman relation are satisfied. 
Furthermore, the leading nonanalytic terms in the chiral expansion of 
 $\langle \overline{q} q \rangle$, $F_\pi$ and $m_\pi$ 
have the correct form, as given 
by chiral perturbation theory at the one-loop level \cite{Gasser}.

We calculate numerically the
contributions of meson loops to $\langle\bar qq\rangle$ 
and $F_\pi $, and find them to be substantial,
of the order of 30\%. 
This is consistent with the $N_{{\rm c}}$-counting scheme.

The outline of the paper is as follows: In Sec.~\ref{sec:formalism}
we recall the model, establish the basic notation and review the
formalism of the effective action. We then proceed to evaluate the
effective action at the quark-loop level (Sec.~\ref{sec:1-q-l})
and at the meson-loop level (Sec.~\ref{sec:loops}), and derive
expressions for the quark condensate  (Sec.~\ref{sec:qq1}), the
meson propagators (Sec.~\ref{sec:K1}), and and the pion decay
constant (Sec.~\ref{sec:fpi}) at the one-meson-loop level.
In Sec.~\ref{sec:chir} we verify explicitly that our expressions have the 
correct chiral expansion. Section Sec.~\ref{sec:res} presents 
our numerical results. Appendices contain technical details 
of our calculation. 

\section{The effective action}
\label{sec:formalism} 

We use the simplest version of the $SU(2)$
Nambu--Jona-Lasinio model \cite{NJL} 
with scalar and pseudoscalar couplings only. The
Lagrangian of the model is given by
\begin{equation}
{\cal L} = \bar q({\rm i}\partial^\mu \gamma_\mu - m)q
 + {\frac{1}{2 a^2}}\left( (\bar q q)^2 + (\bar q{\rm i}\gamma _5 
 \mbox{\boldmath $\tau$} q)^2\right) ,  \label{eq:lagr}
\end{equation}
where the point-like four-quark interaction is characterized by the coupling
constant $1/a^2$ with the dimension of 
inverse energy squared, $q$ stands for up
and down quarks with $N_{{\rm c}}=3$ colors, and $m$ is the current quark
mass. The partition function of the system is given by the path integral
\begin{equation}
Z={\rm e}^{-W}=\int {\cal D}q^{\dagger }{\cal D}q\;{\rm e}^{-%
{\cal I}(q^{\dagger },q)}\,.  \label{eq:partf}
\end{equation}
The integration is over the Grassmann variables $q^{\dagger }$ and $q$, and $%
{\cal I}(q^{\dagger },q)$ is the Euclidean action
\begin{eqnarray}
 {\cal I}(q^{\dagger },q)&=&
 \int d^4xq^{\dagger }\left( \partial _\tau -{\rm i}%
 \mbox{\boldmath $\alpha$} \cdot 
 \mbox{\boldmath $\nabla$} + \beta m\right) q-{\frac{1}{2a^2}}
 \left( (q^{\dagger
 }\beta q)^2+(q^{\dagger }{\rm i}\beta \gamma_5
 \mbox{\boldmath $ \tau$} q)^2\right) . \nonumber \\
& &
\label{eq:euclact}
\end{eqnarray}
The partially bosonized version of the model~\cite{Eguchi:boson} is obtained
by introducing auxiliary meson fields $\Phi :$%
\begin{eqnarray}
Z &=&\int {\cal D}q^{\dagger }{\cal D}q{\cal D}\Phi \exp \Biggl\{ -\int
d^4x\,\biggl[ q^{\dagger }\left( \partial _\tau -{\rm i}
\mbox{\boldmath $\alpha$} \cdot
\mbox{\boldmath $\nabla$} +
 \beta \Gamma_\alpha \Phi_\alpha \right) q  \nonumber \\
&&\ +{\frac 12}a^2\Phi ^2-a^2m\Phi _0 + {\frac 12} a^2 m^2\biggr] 
\Biggr\} ,  \label{eq:Zchi}
\end{eqnarray}
where we use the following notation 
\begin{eqnarray}
\Gamma _0 &=&1;\quad \Gamma_\alpha ={\rm i}\gamma_5\tau_\alpha \; ,  \nonumber \\
\Phi _0 &=&S;\quad \Phi _\alpha =P_\alpha ,\; \alpha =1,2,3\,.  \label{eq:Phi}
\end{eqnarray}
%
The quark
fields, which appear in the exponent of Eq.~(\ref{eq:Zchi}) in a quadratic
form, can be integrated out, and the partition function becomes 
\begin{equation}
Z\equiv {\rm e}^{-W}=\int {\cal D}\Phi {\rm e}^{-{I}(\Phi )},  \label{eq:Z}
\end{equation}
where 
$\int {\cal D}\Phi \equiv \int {\cal D}S
{\cal D}\mbox{\boldmath $P$}$. The so called 
{\em bosonized} Euclidean action $I\left( \Phi \right) $ is given by
\begin{eqnarray}
{I}(\Phi)=-N_{{\rm c}}{\rm Tr}\,\ln D + 
 \int d^4x\, \left ( {\frac{a^2}2}\Phi^2-a^2 m \Phi _0 + 
{\frac{a^2}2} m^2 \right ) .  \label{eq:Seff}
\end{eqnarray}
where we use the Dirac operator in the form
\begin{equation}
D =\partial _\tau + h  \label{eq:D}
\end{equation}
with the one-particle Dirac Hamiltonian $h$ given by 
\begin{equation}
h=-{\rm i}{
\mbox{\boldmath $ \alpha$} \cdot 
\mbox{\boldmath $ \nabla$} }+\beta \Gamma _\alpha \Phi _\alpha
\,.  \label{eq:h}
\end{equation}
The trace in Eq.~(\ref{eq:Seff}) involves an integration over space-time
variables and a matrix trace over the spin and flavor degrees of freedom.
The trace over color gives a factor $N_{{\rm c}}$ which we write explicitly.
Recalling that the quark propagator in the background meson fields $\Phi $
is given by $D^{-1}$, we refer to the term~$N_{{\rm c}}{\rm Tr}\,\ln D$ as
to the {\em quark-loop contribution}.

For simplicity we use the following notation throughout the paper: the
indices $a,b,\dots $ contain the field isospin indices $\alpha,\beta,\dots $
and the space-time coordinates $x_a,x_b,\dots$, {\em i.e.} $a\equiv
\{\alpha,x_a\}$, $b\equiv \{\beta,x_b\}$, {\em etc.} Summation/integration
over repeated indices is understood.

Generally, the quark-loop contribution $N_{{\rm c}}{\rm Tr}\,\ln D$ can have
an imaginary part which is related to the {\em anomalous} terms. In the case
of SU(2) with scalar and pseudoscalar mesons this imaginary part vanishes
identically and the one-quark-loop contribution is given by the real part: $%
N_{{\rm c}}{\rm Tr}\,\ln D={\frac 12}N_{{\rm c}}{\rm Tr}\,\ln (D^{\dagger
}D) $ such that the Euclidean action can also be written in the form 
\begin{equation}
{I}(\Phi )=-{\frac 12}N_{{\rm c}}{\rm Tr}\,\ln (D^{\dagger }D)+
 \int d^4x\, \left ( {\frac{a^2}2}\Phi^2-a^2 m \Phi _0 + 
{\frac{a^2}2} m^2 \right )  .  \label{eq:Seffchi}
\end{equation}
and $D^{\dagger }D$ can be written in the covariant form
\begin{equation}
D^{\dagger }D=\partial _\mu \partial _\mu +i\gamma _\mu \Gamma _\alpha\left(
\partial _\mu \Phi _\alpha\right) + \Phi^2 \;.
\label{DDf}
\end{equation}

The bosonized action ${I}(\Phi )$ is a functional of the fields $\Phi
_\alpha\left( x\right) $.\ The effective action in the one-meson-loop
approximation is given by \cite{Coleman73,Itzykson}: 
\begin{equation}
\Gamma (\Phi )={I}(\Phi )+{\frac 12{\rm Tr}\,\ln \frac{\delta ^2{I}(\Phi )}{%
\delta \Phi \delta \Phi }}\,.  \label{eq:effaction}
\end{equation}

The effective action (\ref{eq:effaction}) has the following 
properties~\cite{Itzykson,Georges}:
\begin{itemize}
\item  At the stationary point of the action, 
the values of the fields $\Phi_a$ represent the 
ground-state expectation values of the field operators.

\item  The inverse field propagators $L^{-1}$ are equal to the second order
derivatives of the effective action: 
\begin{equation}
L^{-1}_{ab} =\frac{\delta ^2\Gamma
\left( \Phi \right) }{\delta \Phi _a\delta \Phi _b}\,,  \label{invprop}
\end{equation}
\end{itemize}
calculated at the stationary point.
The same result can be deduced by a resummation of Feynman graphs, in terms
of quark propagators which are dressed by a general static potential~\cite
{Georges,Georges70}. We refer to the action ${I}(\Phi )$, which is leading
order in $N_c$, as the one-quark-loop approximation, and to the effective
action~(\ref{eq:effaction}), including both the leading and next to leading
order term ${\frac 12{\rm Tr}\,\ln \frac{\delta ^2{I}(\Phi )}{\delta \Phi
\delta \Phi }}$, as the one-meson-loop approximation. We emphasize that in
our approach they are both symmetry conserving approximations.

Most applications of the Nambu Jona-Lasinio model so far, have neglected the
contribution of the second term (the meson loop contribution) to the action (%
\ref{eq:effaction}). The purpose of this paper is to include this
contribution in the description of the vacuum and the pion properties.

\section{The regularized effective action in the one-quark-loop approximation}
\label{sec:1-q-l} 

We first consider the one-quark-loop approximation, in
which the second term of the effective action (\ref{eq:Seffchi}) is neglected.
In this approximation
\begin{eqnarray}
\Gamma (\Phi )&=&I\left( \Phi \right) =-{\frac{N_{{\rm c}}}2}{\rm Tr}\,\ln
(D^{\dagger }D)+\int d^4x \, \left ( 
 {\frac{a^2}2}\Phi^2 - a^2 m \Phi_0 
 + {\frac{a^2}2} m^2 \right ) \;. \nonumber \\
& &  \label{eq:effact}
\end{eqnarray}
The effective action (\ref{eq:effact}) is ultraviolet divergent and requires
regularization. To this aim a fermion-loop cut-off $\Lambda _{{\rm f}}$ is
introduced. Since the model is nonrenormalizable, the cut-off $\Lambda _{%
{\rm f}}$ is kept finite and treated as a parameter. 

In this paper we use two regularization schemes, namely the proper-time
regularization and the covariant four-momentum O(4) regularization, as used
in Ref.~\cite{Jaminon}. We will compare results obtained with these schemes.
The proper-time regularized effective action is given by 
\begin{eqnarray}
I_\Lambda \left( \Phi \right) &=&{\frac{N_{{\rm c}}}2}%
{\rm Tr}\int_{\Lambda _{{\rm f}}^{-2}}^\infty {\frac{ds}s}\,{\rm e}%
^{-sD^{\dagger }D}+
 \int d^4x \, \left ( 
 {\frac{a^2}2}\Phi^2 - a^2 m \Phi_0 
 + {\frac{a^2}2} m^2 \right )\;. \nonumber \\
& &  \label{eq:SeffPT}
\end{eqnarray}
In the O(4) regularization, the four-momentum running in the quark loop is
limited by ${|k|<\Lambda _{{\rm f}}}$ as explained in App.~\ref{app:fregO4}.
In the following we will always use the regularized action and skip the
index $\Lambda $. 
The action~(\ref{eq:SeffPT}) 
has a translationally invariant stationary point at 
 $\Phi = (M_0,0,0,0)$ 
\begin{equation}
\left. {\frac{\delta {I}(\Phi )}{\delta \Phi _a}}\right| _{M_0}=0 \;.
\label{eq:st-point}
\end{equation}
The stationary point $M_0$ is identified with the vacuum 
expectation value of the $S$ field in the one 
quark-loop approximation.
Inserting the regularized
effective action~(\ref{eq:SeffPT}) into Eq.~(\ref{eq:st-point}) we get the
following equation for $M_0$:
\begin{equation}
a^2\left( 1-{\frac m{M_0}}\right) -8N_{{\rm c}}g(M_0)=0\,,  \label{eq:1flGE}
\end{equation}
where the function $g$ is defined in App.~\ref{app:freg}. Equation~(\ref
{eq:1flGE}) is commonly referred to as the gap equation since it determines
the energy gap $2M_0$ between the negative- and positive-energy quark states.
Obviously, $M_0$ plays the role of thge {\em constituent} mass of the quark.

In the one-quark-loop approximation, the inverse meson 
propagators~(\ref{invprop}) are given by the second 
variation of the action with respect to
the fields, taken {\em at the stationary point} $(M_0,0,0,0)$:
\begin{equation}
K^{-1}_{ab} = \left. \frac{\delta ^2I\left( \Phi
\right) }{\delta \Phi_a \delta \Phi_b }\right|_{M_0}.
\end{equation}
In momentum space the inverse 
meson propagators are given by~\cite{Jaminon} 
\begin{equation}
K_{\alpha \beta }^{-1}(q^2)=\delta _{\alpha \beta }\left\{ 4N_{{\rm c}%
}\left( f(M_0,q^2)(q^2+4M_0^2\delta _{\alpha 0})-2g(M_0)\right) +a^2\right\}
,  \label{eq:K0}
\end{equation}
where the function $f$ describing a quark loop with two meson
couplings is given in App.~\ref{app:freg}. Since we are evaluating the
propagator at the stationary point $(M_0,0,0,0)$, 
we can use the gap equation (\ref
{eq:1flGE}) in order to simplify this expression and  obtain the
pseudoscalar and scalar inverse meson propagators in the form
\begin{eqnarray}
K_\pi ^{-1}(q^2) &=&4N_{{\rm c}}f(M_0,q^2)q^2+a^2{\frac m{M_0}},
\label{eq:K0pi} \\
K_\sigma ^{-1}(q^2) &=&4N_{{\rm c}}f(M_0,q^2)(q^2+4M_0^2)+a^2{\frac m{M_0}}.
\label{eq:K0si}
\end{eqnarray}
The propagators have poles at
\begin{eqnarray}
m_\pi ^2 &=&{\frac{a^2m}{4N_{{\rm c}}f(M_0,q^2=-m_\pi ^2)M_0}}\;, \\
m_\sigma ^2 &=&4M_0^2+{\frac{a^2m}{4N_{{\rm c}}f(M_0,q^2=-m_\sigma ^2)M_0}}%
\;,
\end{eqnarray}
where $m_\pi $ and $m_\sigma $ are the on-shell pion and $\sigma $-meson
masses. In the chiral limit $(m\rightarrow 0)$, the pseudoscalar mesons are
massless Goldstone bosons, as expected. The physical quark-meson coupling
constant $g_{\pi qq}$ is given by the residue at the pole of the pion
propagator 
\begin{equation}
g_{\pi qq}^2=\lim_{q^2\to -m_\pi ^2}(q^2+m_\pi ^2){K_\pi }(q^2),
\label{eq:gpiqq-def}
\end{equation}
which gives in the chiral limit 
\begin{equation}
g_{\pi qq}=(4N_{{\rm c}}f(M_0,0))^{-{\frac 12}}\,.  
\label{eq:gpiqq}
\end{equation}

To end this section we recall the $N_{{\rm c}}$-counting rules for quark
and meson loops in the Nambu--Jona-Lasinio model. From the gap equation (\ref
{eq:1flGE}) it follows that the coupling constant $\frac{1}{a^2}$ must be
considered proportional to $\frac{1}{N_{{\rm c}}}$. Each quark-loop
contributes a factor of $N_{{\rm c}}$, which comes from the trace over color
degrees of freedom. Thus, the action ${I}\left( \Phi \right) $ is of
order $N_{{\rm c}}$.
The meson
propagators contain a quark-loop in the denominator and are
hence of order $\frac{1}{N_{{\rm c}}}$, as can be seen directly from Eq.~(%
\ref{eq:K0}). From Eq.~(\ref{eq:gpiqq}) we find that the physical pion-quark
coupling constant $g_{\pi qq}$ is of order $N_{{\rm c}}^{-{\frac 12}}$. The 
physical pion field ($\mbox{\boldmath $\pi$} =
 \mbox{\boldmath $P$}/g_{\pi qq}$) is obtained by
rescaling the meson field 
 $\mbox{\boldmath $P$}$. 
The propagator for 
the physical pion field has the form 
 ${K_\pi(q^2)}/{g_{\pi qq}^2}$, which is 
of order $N_{{\rm c}}^0$. For each quark--physical pion
vertex we have a factor of $g_{\pi qq}\sim N_{{\rm c}}^{-{\frac 12}}$. 
Analogous relations hold for the sigma field. This
way, $N_{{\rm c}}$-counting rules in the NJL model agree 
with those of QCD~\cite{tHooft:nc,Witten:nc}.

\section{The regularized effective action including the 
meson-loop contribution}

\label{sec:loops} In this section we include the meson-loop contribution to
the effective potential, which is the second term of the expression (\ref
{eq:effaction}). This leads to corrections which are next-to-leading order
in $N_{{\rm c}}$. We now consider the effective action in the form (\ref
{eq:effaction}) 
\begin{equation}
\Gamma \left( \Phi \right) =I\left( \Phi \right) +\frac 12{\rm Tr}\,\ln {K}%
^{-1}\left( \Phi \right) \,,  \label{eq:effact1ml}
\end{equation}
where ${K}^{-1}\left( \Phi \right) $ is the functional 
\begin{equation}
K^{-1}_{ab}
= \frac{\delta ^2I\left( \Phi \right) }{\delta \Phi _a
\delta \Phi _b }\,.  \label{eq:K}
\end{equation}

Evaluating the meson-loop term, $\frac 12 {\rm Tr}\,\ln {K}^{-1}\left(\Phi
\right)$, we encounter new divergences which arise due to the integration
over the momentum in the meson-loop. Let us analyze the meson-loop momentum
integrals from a formal point of view. In our model, which has only quark
dynamical degrees of freedom, the meson propagator at the leading $N_{{\rm c}%
}$-level is given by a chain of quark-loops. We have regularized the
quark-loop contribution introducing a fermionic cut-off parameter $\Lambda _{%
{\rm f}}$. This regularization, however, does not restrict the four-momenta
of the mesons. Using the asymptotics of the function 
 $f $ at high momenta it can be easily shown that the leading order field
propagator has the asymptotic expansion $K_{\alpha \beta }^{-1}(q^2) %
\smash{\mathop{\approx}\limits_{q^2\to \infty}} \delta _{\alpha \beta }a^2+%
{\cal O}\left( {\frac 1{q^2}}\right). $ Thus, integrating over the
meson-loop four-momentum in~(\ref{eq:effact1ml}) we encounter a quartic
divergence. Similarly, calculating the pion propagator at the one-meson-loop
level we will obtain both quadratically and logarithmically divergent terms.
We regularize the meson-loop integrals introducing a new {\em bosonic}
cut-off parameter $\Lambda _{{\rm b}}$. We use a covariant O(4)
regularization for the meson loop, consisting in cutting off the meson
four-momenta in the loop integrals at $Q^2=\Lambda _{{\rm b}}^2$. This is
the simplest possible choice.

The one-meson-loop effective action~(\ref{eq:effact1ml}) has a
translationally invariant stationary point 
 $\Phi = (M,0,0,0)$ given by the equation
\begin{equation}
\left. {\frac{\delta \Gamma (\Phi )}{\delta \Phi _a}}\right| _M = 0 \;. 
\label{eq:st-point1}
\end{equation}
The stationary point of the one-meson-loop effective action $\Gamma $ is
denoted by $M$, in order to distinguish it from the stationary point $M_0$
of the quark-loop action defined in Eq.~(\ref{eq:st-point}). 

\begin{figure}[tbp]
\xslide{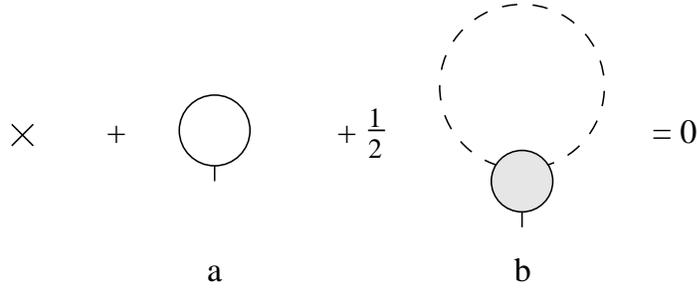}{5cm}{0}{280}{610}{500}
\caption{Diagramatic representation of 
Eq.~(\protect{\ref
{eq:GE1}}):  (a)  the quark-loop contribution, 
(b) the meson-loop contribution. The cross denotes the constant 
term $a^2 (1 - m/M)$. The short external lines 
denote the scalar-isoscalar coupling, 
the solid line in (a) denotes the (regularized) quark propagator,
the dashed line in (b) denotes the meson propagator 
 $\widetilde{K}$ defined in Eq.~\protect{\ref{eq:K0iSCS}},
and the filled blob denotes the
meson vertex $S_{0bc}$, explicitly given
in App.~\protect{\ref{app:Seff-exp}}.}
\label{fig:GE1}
\end{figure}

We note that $M$ {\em does not} correspond to the pole
of the quark propagator at the one-meson-loop level --- it is not the quark
mass. It is only the local (Hartree) 
contribution to the quark propagator mass.
The quantity $M$ is the expectation value of the $S$ field
when meson-loop effects included. 
We may rewrite Eq.~(\ref{eq:st-point1}) in the form
\begin{eqnarray}
\left. {\frac{\delta \Gamma (\Phi )}{\delta \Phi _a}}\right| _M=S_a(M)+{%
\frac 12}S_{abc}(M)K_{bc}(M)=0\,,  \label{eq:GE1-S}
\end{eqnarray}
where we have introduced the following short-hand notation for the $n$-leg
quark-loop meson vertices 
\begin{equation}
S_{a_1a_2\dots a_n}(M)\equiv \left. {\frac{\delta ^n{I}(\Phi )}{\delta \Phi
_{a_1}\delta \Phi _{a_2}\dots \delta \Phi _{a_n}}}\right| _M\,.
\label{eq:Sabc}
\end{equation}
For $a=1,2,3$, the condition (\ref{eq:GE1-S}) is trivially satisfied with a
vanishing pion field. From the variation over the sigma field we obtain 
\begin{equation}
S_0(M)+{\frac 12}S_{0bc}(M)K_{bc}(M)=0\,.  \label{eq:GE1}
\end{equation}
We represent Eq.~(\ref{eq:GE1}) digramatically in Fig.~\ref{fig:GE1}.
The first two terms where present at the quark-loop level. The cross
is the contact term contribution to $S_0(M)$, and (a) denotes the 
quark-loop contribution. The term  (b) is the meson-loop
contribution.
The meaning of symbols is explained in the caption.
Using the $N_{{\rm c}}$-counting rules (see Section \ref{sec:1-q-l}) it can
be verified that the meson-loop diagram (b) is of order $N_{{\rm c}}^0$ and
hence suppressed by one power in $\frac 1{N_{{\rm c}}}$ compared to the
quark-loop diagram, which is of order $N_{{\rm c}}$.
The leading-order meson propagators in Eq.~(\ref{eq:GE1}) $K_{bc}(M)$ are
evaluated at the stationary point 
of the one-meson-loop effective action, $M$. In 
momentum space, similarly to Eq.~(\ref{eq:K0}), we obtain 
\begin{eqnarray}
K_{\alpha \beta }^{-1}(M,q^2) &=&\delta _{\alpha \beta }\left\{ 4N_{{\rm c}%
}\left( f(M,q^2)(q^2+4M^2\delta _{\alpha 0})-2g(M)\right) +a^2\right\}  
\nonumber \\
&=&\widetilde{K}_{\alpha \beta }^{-1}(M,q^2)+\delta _{\alpha \beta }\Delta
(M)\,,  \label{eq:K0iS}
\end{eqnarray}
where we have defined
\begin{equation}
\widetilde{K}_{\alpha \beta }^{-1}(M,q^2)=
\delta _{\alpha \beta } \left [ 4N_{{\rm c}} f(M,q^2)
(q^2+4M^2\delta _{\alpha 0}) + \frac{a^2 m}{M} \right ]  \label{eq:K0iSCS}
\end{equation}
and 
\begin{equation}
\Delta (M)=-8N_{{\rm c}}g(M)+a^2 (1 - \frac{m}{M}).  \label{eq:dK0iS}
\end{equation}
Since $\Delta (M)$ is exactly equal to the first term $S_0(M)$ in 
Eq.~(\ref{eq:GE1}) we have 
\begin{equation}
\Delta (M)=S_0(M)=-{\frac 12}S_{0bc}(M)K_{bc}(M)={\cal O}\left( N_{{\rm c}%
}^0\right) \,,
\end{equation}
which shows that $\Delta (M)$ is one order in $N_{{\rm c}}$ suppressed
compared to the full propagators~(\ref{eq:K0iS}) which are of order $N_{{\rm %
c}}$. Inserting the decomposition~(\ref{eq:K0iS}) into Eq.~(\ref
{eq:GE1}) we obtain
\begin{eqnarray}
& & S_0(M) + {\frac 12}S_{0bc}(M)\widetilde{K}_{bc}(M)-{\frac 12}S_{0bc}(M)%
\widetilde{K}_{bd}(M)S_0(M)\widetilde{K}_{dc}(M)+\dots =0\;. \nonumber \\
& &
\label{eq:GE1-exp}
\end{eqnarray}
The third term in this expansion is of order $\frac 1{N_{{\rm c}}}$ and
hence one order in $N_{{\rm c}}$ suppressed compared to the meson-loop
contribution (the second term). Further terms are suppressed by even higher
powers of $N_{{\rm c}}$. At the present level of approximation 
we should 
therefore keep only the first two terms in 
Eq.~(\ref{eq:GE1-exp}) --- otherwise consistency in $N_c$-counting
is lost.  

The explicit form of Eq.~(\ref{eq:GE1-exp}) used in 
numerical calculations has the form
\begin{eqnarray}
& & a^2 \left(1 - \frac{m}{M} \right) - 8 N_c \, g(M) + 
 \frac{N_c}{4 \pi^4} \int d^4 Q \, \times \nonumber \\
& & \left\{ \left [2 f(M,0) + \frac{d}{dM^2}\left (f(M,Q^2)(Q^2+M^2)
\right ) \right]
 \widetilde{K}_\sigma(M,Q^2) \right. \nonumber \\
& & + \left. 3 \left [2 f(M,0) + \frac{d}{dM^2} f(M,Q^2) Q^2 \right]  
 \widetilde{K}_\pi(M,Q^2) \right\} = 0 \;, 
\label{eq:GE1-f}
\end{eqnarray}

\section{Quark condensate}
\label{sec:qq1} 

The quark condensate $\langle \bar qq\rangle $ is given by 
 \mbox{$\langle \bar qq\rangle = {\delta \Gamma (\Phi )}/{\delta m}$} 
which from Eq.~(\ref{eq:effact1ml}) immediately gives
\begin{equation}
\langle \bar qq\rangle =-a^2(M-m)\,.  \label{eq:qq1}
\end{equation}
Comparing the result (\ref{eq:qq1}) with the
corresponding expression at the one-quark-loop level, 
 \mbox{$\langle \bar qq\rangle =-a^2(M_0-m)$}, 
we can see that it is analogous, with  $M_0$
replaced by $M$. We shall compare our theoretical results to the empirical
value~\cite{Narison90}
\begin{equation}
\langle \bar uu\rangle =\langle \bar dd\rangle ={\frac 12}\langle \bar
qq\rangle =-((250\pm 50)\ {\rm MeV})^3\,.  \label{eq:qq-exp}
\end{equation}

\section{Meson propagators}
\label{sec:K1} 

\begin{figure}[tbp]
\xslide{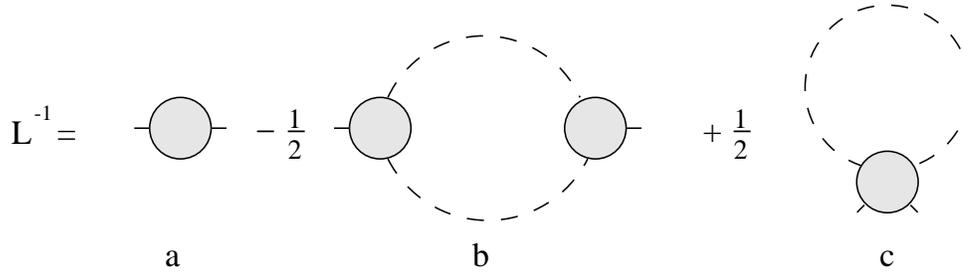}{5cm}{0}{280}{610}{500}
\caption{Diagrams contributing to the inverse meson propagator in
one-meson-loop approximation: (a) quark-loop contribution, 
(b--c) meson-loop contributions. The dashed line denotes the meson
propagator $\widetilde K$, and the filled quark-loop meson vertices
are defined in App.~{\protect\ref{app:Seff-exp}}.}
\label{fig:K1i}
\end{figure}

The inverse meson propagators, including the one-meson-loop
effects, are equal to the second order variation (\ref{invprop}) of the
effective action~(\ref{eq:effact1ml}) with respect to the fields taken at
the stationary point $M$:
\begin{equation}
L^{-1}_{ab} =\left. \frac{\delta
^2\Gamma \left( \Phi \right) }{\delta \Phi _a \delta \Phi
_b }\right| _M\,.  \label{eq:K1Phi}
\end{equation}
Performing the variation in~(\ref{eq:K1Phi}) we can express the propagators
in one-meson-loop approximation in terms of the 
functional $K$ (\ref{eq:K}) and the quark-loop meson vertices~(\ref{eq:Sabc}): 
\begin{eqnarray}
L_{ab}^{-1} &=&K_{ab}^{-1}(M)-{\frac 12}%
S_{ace}(M)K_{cd}(M)K_{ef}(M)S_{dfb}(M)  \nonumber \\
&&\ +{\frac 12}S_{abcd}(M)K_{cd}(M)\,.  \label{eq:K1i}
\end{eqnarray}
The diagrammatic representation of this equation is shown 
in Fig.~\ref{fig:K1i}. The first term is the leading-order 
(quark-loop) contribution 
 ${K}^{-1}(M)$. We have to bare in mind, however, that here it differs from
the inverse propagator ${K}^{-1}(M_0)$ in the one-quark-loop approximation, 
because it is
evaluated at the new stationary point $M$ of the one-meson-loop effective
action. This means that the quark propagators appearing in the meson
propagator $K(M)$ are dressed by solving Eq.(\ref{eq:GE1-exp}). 
The second (b) and third term (c) are the
one-meson-loop contributions to the inverse meson propagator. The dashed
lines in the loops correspond to a pion or to a sigma meson. The 
one-meson-loop contributions are suppressed one order in $N_{{\rm c}}$, which
can be easily shown using the $N_{{\rm c}}$-counting rules.

In Eq.~(\ref{eq:K1i}) $M$ is the solution of the Eq.~(\ref
{eq:GE1-exp}), in which only the first two terms are retained, so as to be
consistent with $N_{{\rm c}}$-counting. It is crucial to maintain the same 
$N_{{\rm c}}$-counting consistency in the calculation of the meson
propagators~(\ref{eq:K1i}). 
We include in the inverse propagator $L^{-1}$ contributions
which are of order $N_{{\rm c}}$ and $N_{{\rm c}}^0$. In the first term $%
K^{-1}$, which is defined in Eq.~(\ref{eq:K0iS}), we keep both the
contributions $\widetilde{K}^{-1}$, of order $N_{{\rm c}}$, and the
contribution $\Delta (M)$, of order $N_{{\rm c}}^0$. In the remaining
(meson-loop) terms of Eq.~(\ref{eq:K1i}), the vertices $S$ are of order $N_{%
{\rm c}}$ (because they contain a quark loop) and the leading order of the $K
$ operators is $N_{{\rm c}}^{-1}$. Therefore $N_{{\rm c}}$-counting
consistency requires the substitution of $K$ with $\widetilde{K}$
in the two meson-loop terms. Keeping all terms to order $N_{{\rm c}%
}^0$ we obtain the inverse meson propagators in the form
\begin{eqnarray}
L_{ab}^{-1} &=&K_{ab}^{-1}(M)-{\frac 12}S_{ace}(M)\widetilde{K}_{cd}(M)%
\widetilde{K}_{ef}(M)S_{dfb}(M)  \nonumber \\
&&+{\frac 12}S_{abcd}(M)\widetilde{K}_{cd}(M),  \label{eq:K1ii}
\end{eqnarray}
where the propagators $\widetilde{K}_{ab}$ in the meson loops are given by~(%
\ref{eq:K0iSCS}). 

Despite the simple form~(\ref{eq:K1ii}) of inverse meson
propagators, their
numerical evaluation for an arbitrary momentum is quite involved. It requires
integration over the meson-loop four-momentum $Q$, over the momenta in the
quark-loops as well as over the proper-time (or Feynman) parameters. Even
after taking some of the integrals analytically we still end up with
many-dimensional integrals that must be evaluated numerically. However, we
are first of all interested in pionic properties. Since the pion is very
light, we can determine the pion mass and the pion-quark coupling constant
from the low-momentum expansion of the pion propagator. This leads to a
radical simplification of the numerical calculation.
The inverse pion propagator in momentum space can be expanded in the form 
\begin{equation}
L_{\pi}^{-1}(q^2) = L_{\pi}^{-1}(0) + Z_\pi (0) q^2 + {\cal O}(q^4)\, .
\label{eq:K1i-exp}
\end{equation}
Both the constant term, $L_{\pi}^{-1}(0)$, and the $q^2$-term, $Z_\pi(0)$,
can be expressed as a sum of three contributions corresponding to the
diagrams in Fig.~\ref{fig:K1i}. 

We will now check that the pion remains massless when 
the $\frac{1}{N_{{\rm c}}}$ meson-loop corrections 
are taken into account. 
It is crucial for the proof that 
the expectation value of the $S$ field, $M$, is 
determined from the stationary-point condition for the
effective action. 
We note that at the one-meson-loop level the stationary
point of the effective action and the quark self energy do not coincide. 
In
the approaches of Refs.~\cite{Cao92,Domitrovich93,Heid:loops} where the gap
equation at the one-meson-loop level includes the quark self-energy diagram
the Goldstone theorem does not hold. It does 
hold in the work of Ref.~\cite{Tegen95}, where the quark 
self energy diagram has been replaced by a
tadpole diagram. Thus contributions of the same order in $N_{{\rm c}}$ as
the calculated meson-loop correction have been excluded in order to fulfill
the Goldstone theorem. The retained tadpole diagram is exactly the one
obtained by dressing the quark propagator by a general static potential and
hence in the approach of Ref.~\cite{Tegen95} one can recognize a symmetry
conserving approximation~\cite{Georges}.

In order to check the validity of the Goldstone theorem we have to evaluate
the leading-order term $L_{\pi}^{-1}(0)$ in the low-momentum expansion of
the inverse pion propagator. We write the inverse pion propagator at zero
momentum as a sum of the contributions of the three diagrams in Fig.~\ref
{fig:K1i}, in which we have now couplings with zero external pion momentum.
This leads to a significant simplification in the evaluation of the
diagrams. The algebra is straightforward but tedious, 
and we refer the reader to
Ref.~\cite{emil:phd} for details.
The result is
\begin{eqnarray}
L_\pi^{-1}(0) &=& a^2-8N_{{\rm c}}g(M) +
 \frac{N_c}{4 \pi^4} \int d^4 Q \, \times \nonumber \\
& & \left\{ \left [2 f(M,0) + \frac{d}{dM^2}\left (f(M,Q^2)(Q^2+M^2)
\right ) \right]
 \widetilde{K}_\sigma(M,Q^2) \right. \nonumber \\
& & + \left. 3 \left [2 f(M,0) + \frac{d}{dM^2} f(M,Q^2) Q^2 \right]  
 \widetilde{K}_\pi(M,Q^2) \right\}\;, 
\label{eq:K10}
\end{eqnarray}
Using Eq.~(\ref{eq:GE1-f}) in (\ref{eq:K10})  we
obtain immediately
\begin{equation}
L_\pi^{-1}(0)=a^2{\frac m{M}},  \label{eq:K10-f}
\end{equation}
which is analogous to the one-quark-loop result with $M_0$ replaced with $M$.
In the chiral limit ($m=0$) we have $L_\pi^{-1}(0)=0$, showing that the
pions are massless. Thus, as expected, 
the Goldstone theorem is satisfied in the one meson-loop approximation.
This is a general feature. If we derive both the gap equation and the meson
propagators by performing functional derivatives of the effective action, we
have a symmetry-conserving approximation. The Goldstone pion reflects this.

For non-vanishing quark current mass $m$ the chiral symmetry is explicitly
broken and the pion acquires a finite mass. It is given by the position of
the pole of the pion propagator. Using the inverse pion propagator we obtain 
\begin{equation}
L_\pi^{-1}(-m_\pi ^2) = L_\pi^{-1}(0)-Z_\pi(-m_\pi^2)\, m_\pi^2=0
\label{eq:K1iPi-g}
\end{equation}
and using Eq.~(\ref{eq:K10-f}) we obtain the on-shell pion mass as a
solution of the following equation 
\begin{equation}
\label{eq:shpm}
m_\pi^2 = {\frac{a^2\, m }{Z_\pi(-m_\pi^2)\, M}} \, .
\end{equation}
Since $m_\pi$ is very small we can approximate $Z_\pi(-m_\pi^2)$ by its
value at zero momentum. Then we can calculate the pion mass in one
meson-loop approximation using the results for the low-momentum expansion of
the pion propagator 
\begin{equation}
m_\pi^2 = {\frac{a^2\, m }{Z_\pi(0)\, M}} + {\cal O}(m_\pi^4) \, .
\label{eq:mpi1}
\end{equation}

The pion-quark coupling constant can be determined on the one-meson-loop
level similarly to~(\ref{eq:gpiqq-def}) as the residue of the pion
propagator at its pole 
\begin{equation}
g_{\pi qq}^{2}=\lim_{q^2\to -m_\pi ^2}(q^2+m_\pi^2) L_\pi^{-1}(q^2).
\label{eq:gpiqq-def1}
\end{equation}
In the chiral limit we obtain the simple result 
\begin{equation}
g_{\pi qq}=(Z_\pi(0))^{-{\frac 12}}.  \label{eq:gpiqq1}
\end{equation}

\begin{figure}
\xslide{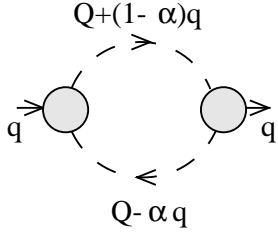}{5cm}{0}{300}{605}{480}
\caption{The routing ambiguity in the 
meson-loop contributions of 
Fig.~\protect{\ref{fig:K1i}}, diagram (b), 
to the inverse pion propagator.} 
\label{fig:K1ibw}
\end{figure}

In order to evaluate $m_\pi$ and $g_{\pi qq}$ we need, 
in addition to $M$ which we obtain by solving Eq.~(\ref{eq:GE1-exp}), 
the wave-function renormalization $Z_\pi(0)$. 
The steps in this calculation are as follows: We 
expand  
various contributions of Fig.~\ref{fig:K1i} in powers of the external
momentum $q$ and isolate the coefficient of $q^2$. The momentum
in loops is cut by the mesonic cut-off $\Lambda_b$. In the case 
of diagram (b) we have, for this cut-off prescription, an
ambiguity in routing the momentum in the loop, 
as shown in Fig.~\ref{fig:K1ibw}.   
Depending on the value of the routing parameter $\alpha$, we get 
different numerical results. In this paper we average with equal
weights over all $\alpha$ from 0 to 1. This prescription is arbitrary,
and one could use another one, {\em e.g.} choose $\alpha = \half$.
As we will see, the choice of routing does not affect fundamental
properties, such as GMOR relation, or the form of the leading nonanalytic
terms in the chiral expansion (see Sec. \ref{sec:chir}). Obviously, it does
affect the numerical value of $m_\pi$ and $F_\pi$ to some extent.
The algebra and final expressions for $Z_\pi(0)$ are lengthy, and the 
reader is referred to Ref.~\cite{emil:phd} for details.

\section{The pion decay constant}
\label{sec:fpi} 
The pion decay constant is an 
important quantity in our
calculations, since we use its empirical value in order to fix
the parameters of the model. Its evaluation allows us also to check
explicitly the Goldberger-Treiman relation for the quarks, which gives the
connection between $M$, the pion-quark coupling
constant in the chiral limit. 
The Gell-Mann--Oakes--Renner relation
then follows straightforwardly.

The pion decay constant $F_\pi$ is defined by the matrix element for the
weak pion decay 
\begin{equation}
\langle 0 | A^\alpha_\mu (z) | \pi^\beta (q) \rangle = -{\rm i} 
\delta^{\alpha\beta} q_\mu F_\pi {\rm e}^{-{\rm i} q\cdot z} \, ,  
\label{eq:fpi-def}
\end{equation}
where $A^\alpha_\mu = \bar q \gamma_\mu\gamma_5\tau^\alpha/2 q$ 
is the axial current and 
 $\mbox{\boldmath $\pi$}$ is the physical pion
field $\mbox{\boldmath $\pi$} = \mbox{\boldmath $P$} / g_{\pi qq}$. 
Applying the
Lehmann-Symanzik-Zimmermann reduction formula we can relate the pion decay
constant $F_\pi=F(-m_\pi^2)$ to the form factor $F(q^2)$, which can be
expressed in terms of the expectation value of the time-ordered product 
\begin{equation}
\langle 0 | T[ A^\alpha_\mu (z) , \pi^\beta(0)] | 0 \rangle = -{\rm i} 
\delta^{\alpha\beta}\int {\frac{d^4 q }{(2\pi)^4}} F(q^2) {\frac{q_\mu
}{q^2 + m_\pi^2}} {\rm e}%
^{-{\rm i} q\cdot z} \, .  \label{eq:wpd}
\end{equation}
To evaluate this matrix element we couple an external axial source 
$j_\mu^\alpha$
in the effective action, which gives for the Dirac operator ({\em cf.}\ Eq.~(%
\ref{eq:D})) 
\begin{equation}
D = \beta \left(-i\gamma_\mu\partial_\mu+ \Gamma_\beta \Phi_\beta + 
\gamma^\mu \gamma_5 {\frac{\tau_\alpha }{2}} j^\alpha_\mu\right)\,.
\end{equation}
Inserting this operator in the action we obtain the generating functional in
the presence of the source $j^\alpha_\mu$ for the external axial field. Then 
the
matrix element~(\ref{eq:wpd}) can be expressed as a variation of the
generating functional 
\begin{equation}
\langle 0 | T[A^\alpha_\mu (z) , \pi^\beta(0)]| 0 \rangle = \frac1{g_{\pi qq}} 
\left. \int d^4x {\delta^2 \Gamma\over \delta j^\alpha_\mu(z)\delta \pi^\beta 
(x)}\right|_{j=0} L_\pi(x)\,. 
\label{eq:dWdj}
\end{equation}
The rhs of this equation has the same structure as the one-meson-loop pion
propagator~(\ref{eq:K1Phi}) with one derivative with respect to the meson
field replaced by a derivative with respect to the axial source. Hence, in
the one-meson-loop approximation, we obtain contributions to the pion decay
constant from diagrams analogous to those contributing to the inverse meson
propagators (Fig.~\ref{fig:K1i}), but with one of the external meson
couplings replaced by an axial coupling.

We calculate $F_\pi$ from the low-momentum expansion as we did for the one
meson-loop pion propagator, with the same routing 
prescription. The difference is that now we pick up the terms
linear in the external momentum $q_\mu$. Also,
due to the axial coupling, we obtain different spin-isospin structure of the
quark-loop functions. The calculation is straightforward 
but tedious, and details are in Ref.~\cite{emil:phd}.
The final result is 
\begin{equation}
F_\pi = g_{\pi qq}M Z_\pi(m_\pi^2)\, .  \label{eq:GT10}
\end{equation}
In the chiral limit this reduces to the Goldberger-Treiman relation \cite{GT}
for the quarks 
\begin{equation}
g_{\pi qq} \, F_\pi = M\, .  \label{eq:GT1}
\end{equation}
Using this result and the expressions for the pion mass~(\ref{eq:mpi1}) and
the quark condensate~(\ref{eq:qq1}) at the one-meson-loop level we recover
also the Gell-Mann--Oakes--Renner relation~\cite{GMOR}:
\begin{equation}
m \langle \bar q q \rangle = - m_\pi^2 \, F_\pi^2 + {\cal O} (m_\pi^4)\, .
\label{eq:GMOR1}
\end{equation}

We have shown that in our effective action approach the Goldstone theorem,
the Goldberger-Treiman, and the Gell-Mann--Oakes--Renner relations are valid
with the meson loops included. Thus the basic relations following from
Ward-Takahashi identities are satisfied in the one-meson-loop approximation.
Again, this is a feature of a symmetry-conserving approximation.

\section{Chiral expansions}
\label{sec:chir}

In this section we check that the leading nonanalytic 
terms in chiral expansion of the quark condensate, the pion mass
and the pion decay constant agree with 
the one-loop results of chiral perturbation theory. 

In chiral perturbation theory \cite{Gasser} we have
\begin{equation}
\label{eq:gl}
\langle \overline{q} q \rangle = \langle \overline{q} q \rangle_0 
\left ( 1 - \frac{3 m_\pi^2}{32 \pi^2 F_\pi^2} 
\log m_\pi^2 + ... \right ) ,
\end{equation}
where $\langle \overline{q} q \rangle_0$ is the value of the quark 
condensate in the chiral limit. 
We wish to show the same result holds in the NJL model with meson loops.
Equation (\ref{eq:GE1-f}), which we may write as 
 $G(M,m) = 0$, defines $M$ as an implicit function of $m$.
We have therefore 
\begin{equation} 
 \frac{dM}{dm} = 
   - \frac{\frac{\partial G}{\partial m}}{\frac{\partial G}{\partial M}} \;.
\label{eq:implicit}
\end{equation}
We find explicitly
\begin{equation}
\label{eq:dM}
\!\!\!\!\!\!\!\!\! 
\frac{\partial G}{\partial M} = a^2 m / M^2 - 8 N_c dg(M)/dM + 
 {\cal O}(N_c^0) = \frac{4 M}{g_{\pi qq}^2} + {\cal O}(N_c^0) + {\cal O}(m)\;,
\end{equation}
where we have used the identity $dg(M)/dM = - 2 M f(M,0)$, 
and the relation (\ref{eq:gpiqq}).
For $\frac{\partial G}{\partial m}$
the only infrared nonanalytic term is the pion-loop term, the last one in
Eq.~(\ref{eq:GE1-f}). Close to the pion pole
this piece is 
\begin{equation}
\label{eq:pionpole}
3 N_c \frac{1}{4 \pi^2} \int_0^{\Lambda_b} d(Q^2) Q^2 
\frac{2 f(M,0)}{4N_c f(M,0) Q^2 + a^2 m/M},  
\end{equation}
which yields the nonanalytic term 
 $\frac{3}{8 \pi^2} m_\pi^2 \log(m_\pi^2/\Lambda^2)$ via Eq.~(\ref{eq:mpi1}).
From  (\ref{eq:dM}) and (\ref{eq:implicit}) we find immediately
\begin{equation}
\label{eq:n}
\frac{dM}{dm_\pi^2} = - \frac{3 g_{\pi qq}^2}{32 \pi^2 M} 
\frac{d}{dm_\pi^2 } \left ( m_\pi^2 \log(m_\pi^2/\Lambda^2) \right ) 
+ {\cal O}(1/N_c^2) + {\cal O}(m_\pi^2). 
\end{equation}
Using Eq.~(\ref{eq:GT1}) in (\ref{eq:n}) 
we get
\begin{equation}
M = M_{m=0}
\left ( 1 - \frac{3 m_\pi^2}{32 \pi^2 F_\pi^2} 
\log m_\pi^2 + ... \right ) \;,
\label{chi:M}
\end{equation}
where $M_{m=0}$ is the value of $M$ in the chiral limit.
From the fact that
 $\langle \overline{q} q \rangle = - a^2 (M-m)$ we 
immediately establish the 
desired result (\ref{eq:gl}).
Note, that for $m_\pi$ and $F_\pi$ in the above expression 
we use the leading-$N_c$ pieces.
The subleading pieces to these quantities
are relevant in relative $1/N_c^2$ corrections
to  $\langle \overline{q} q \rangle$, which is not considered
at the present level of approximation. 

Chiral perturbation theory also determines the leading nonanalytic
terms in the expansion of $m_\pi$ and $F_\pi$. 
Denoting the leading-$N_c$ values of the pion mass and decay constant by
 $m_{\pi,0}$ and $F_{\pi,0}$ (in the notation of Ref.~\cite{Gasser} these
are $M$ and $F$ --- the tree-level pion mass and decay 
constant), we have
\begin{equation}
m_\pi^2 = m_{\pi,0}^2 \left ( 
 1 + \frac{m_{\pi,0}^2}{32 \pi^2 F_{\pi,0}^2} \log m_{\pi,0}^2 + ... 
 \right ) \; 
\label{chi:m}
\end{equation}
\begin{equation}
F_\pi = F_{\pi,0} \left ( 
 1 - \frac{m_{\pi,0}^2}{16 \pi^2 F_{\pi,0}^2} \log m_{\pi,0}^2 + ... 
 \right ) .\; 
\label{chi:F}
\end{equation}
In order to check that these expansions hold in our treatment of the NJL model,
we first look at the chiral expansion of $Z_\pi(0)$. There are cancellations
occuring between the quark and meson loop contributions, and the final
answer is 
\begin{equation}
Z_\pi(0) = Z_{\pi,0}(0) \left ( 
 1 + \frac{m_{\pi,0}^2}{16 \pi^2 F_{\pi,0}^2} \log m_{\pi,0}^2 + ... 
 \right ) .\; 
\label{chi:Z}
\end{equation}
The above result is independent of the routing ambiguity of
Fig.~\ref{fig:K1ibw}. Using expansions (\ref{chi:M}) and (\ref{chi:Z})
in Eqs.~(\ref{eq:mpi1}) and (\ref{eq:GT1}) we obtain the desired 
expressions (\ref{chi:m}-\ref{chi:F}).

We emphasize that the agreement of the results of this 
section with chiral perturbation theory is a direct consequence of 
the symmetry-conserving approximation.

We have also performed a numerical study in order to check how
good is expansion (\ref{eq:gl}) in our case. For typical parameters
($a = 162 {\rm MeV}$, $\Lambda_f = \Lambda_b = 755 {\rm MeV}$, $m = 0$) 
we find that $\langle \overline{q} q \rangle = -(173.8{\rm MeV})^3$, and for 
same parameter but $m = 13{\rm MeV}$, which fits the pion mass to its
physical value, we have 
 $\langle \overline{q} q \rangle = -(175.0{\rm MeV})^3$, {\em i.e.}
a 5\% 
change. In comparison, Eq.~(\ref{eq:gl}) would give 8\%
with the chiral scale under the $\log$ taken to be $1 {\rm GeV}$. This 
shows that the leading nonanalytic correction accounts for
a major part of the
change of the condensate as one departs from the chiral limit.

\section{Numerical results and discussion}

\label{sec:res} At the one-quark-loop level the model has the following
parameters: the quark-quark coupling constant $\frac 1{a^2}$, the
quark cut-off $\Lambda _{{\rm f}}$, and the current quark mass $m$. With
the meson loop included we have in addition the meson cut-off $\Lambda _{%
{\rm b}}$.
We use the empirical values for the observables in the meson sector, namely $%
F_\pi=93$~MeV, and $m_\pi=139$~MeV in order to fix the model parameters.
After fitting the pion mass and decay constant, we are left with two free
parameters. We could fix one of them by fitting the phenomenological value
of the quark condensate $\langle \bar qq\rangle $. The results of our
calculation for $\langle \bar qq\rangle $, however, are quite sensitive to
the cut-off procedure. Furthermore, the experimental range for 
 $\langle \bar qq\rangle$ is very wide, 
 $(-{\frac 12}\langle \bar qq\rangle )_{{\rm exp}}^{\frac 13}=250\pm 50$~MeV. 
For these reasons we do not use the value of
the quark condensate to fix the parameters. Since we do not know the
particular physics underlying the regularization of the theory, QCD cannot
be a guide to fix the loop cut-offs $\Lambda _{{\rm f}}$ and $\Lambda _{{\rm %
b}}$. In this exploratory calculation we display results obtained for four
values of the ratio $\Lambda _{{\rm b}}/\Lambda _{{\rm f}} = 0$, $0.5$, $1$,
and $1.5$. Using the gap equation we eliminate the coupling constant 
$1/a^2$ in
favor $M$, which we treat as a free parameter.
Thus all quantities are presented as a function of $M$ for different values
of the ratio $\Lambda_{{\rm b}}/\Lambda_{{\rm f}}$.

\begin{figure}[tbp]
\xslide{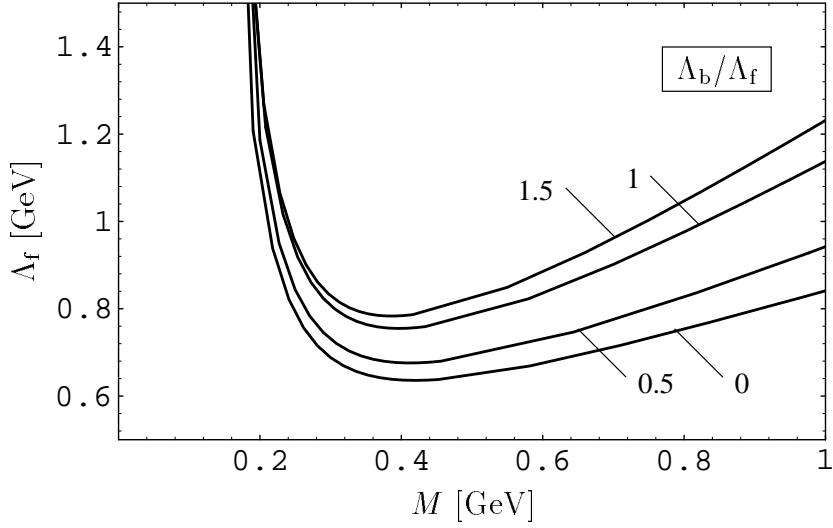}{7.5cm}{80}{280}{500}{490}
\caption{The fermionic cut-off $\Lambda_{{\rm f}}$ for proper-time quark-loop
regularization as a function of $M$  for
different values of $\Lambda_{{\rm b}} / \Lambda_{{\rm f}} $. The curves
correspond to  ${\Lambda_{{\rm b}}/ \Lambda_{{\rm f}}}= $ 0, 0.5, 1, 1.5 .}
\label{fig:MLfPT}
\end{figure}
\begin{figure}[tbp]
\xslide{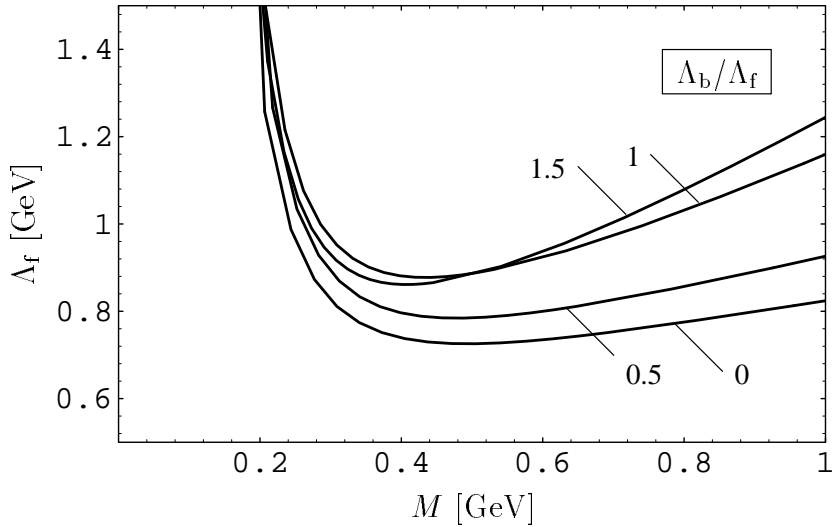}{7.5cm}{80}{280}{500}{490}
\caption{The fermionic cut-off $\Lambda_{{\rm f}}$ for O(4) quark-loop
regularization as a function of $M$  for
different values of $\Lambda_{{\rm b}} / \Lambda_{{\rm f}} $. The curves
correspond to  ${\Lambda_{{\rm b}}/ \Lambda_{{\rm f}}}= $ 0, 0.5, 1, 1.5 .
Note that the
curve for ${\Lambda_{{\rm b}}/ \Lambda_{{\rm f}} } = $ 1.5 has the critical $%
\Lambda_{{\rm f}}$ smaller than for ${\Lambda_{{\rm b}}/ \Lambda_{{\rm f}} }
= $ 1.}
\label{fig:MLfO4}
\end{figure}

We present the calculations performed in the chiral limit ($m=0$).
Introducing a nonzero current quark mass $m$ leads to small corrections to
the calculated quantities. Since at this point we are mainly interested in
the general behavior of the quark condensate
as a function of the model parameters we find it useful to start by
considering the chiral limit.
The quantity $M$ in one-meson-loop approximation is given by
the solution of Eq.~(\ref{eq:GE1}). We use this equation in
order to determine the fermionic cut-off $\Lambda_{{\rm f}}$ for given
values of $M$ and $\Lambda_{{\rm b}}/ \Lambda_{{\rm f}}$. We plot $\Lambda_{%
{\rm f}}$ as a function of $M$ for different
values of the ratio $\Lambda _{{\rm b}}/\Lambda _{{\rm f}}$ of the bosonic
to the fermionic cut-off. The results in the chiral limit ($m=0$) with
proper-time and O(4) fermion-loop regularization are presented in Figs.~\ref
{fig:MLfPT} and \ref{fig:MLfO4}, respectively. The pion decay constant $%
F_\pi $ is fitted to reproduce the experimental value. For $\Lambda _{{\rm b}%
}=0$ we recover the results of the one-quark-loop approximation, so that
the difference between the lowest curve and the other curves is a measure of
the meson loop contribution. For $\Lambda_{{\rm f}}$ lower than a critical
value we do not have spontaneous breaking of the chiral symmetry and 
Eq.~(\ref{eq:1flGE}) has no solution. With increasing mesonic cut-off $%
\Lambda _{{\rm b}}$ the critical value of $\Lambda _{{\rm f}}$ increases and
the curve ``moves'' up until $\Lambda _{{\rm b}}/\Lambda _{{\rm f}}$ reaches 
$\sim 1.5$ for the proper-time cut-off and $\sim 1.2$ for the O(4) one. For
larger values of $\Lambda _{{\rm b}}/\Lambda _{{\rm f}}$ the minimal value
of $\Lambda _{{\rm f}}$ starts decreasing and the curve moves down with
increasing $\Lambda _{{\rm b}}/\Lambda _{{\rm f}}$. This differs from the
results of Dmitra\v sinovi\' c {\em at al.}~\cite{Tegen95} with
Pauli-Villars fermion-loop regularization where the curves keep moving
towards larger values of $\Lambda _{{\rm f}}$ as $\Lambda _{{\rm b}}/\Lambda
_{{\rm f}}$ increases (see Fig.~11 in~\cite{Tegen95}). For large mesonic
cut-offs, $\Lambda _{{\rm b}}/\Lambda _{{\rm f}}$ over 2 for the proper-time
fermion regularization and $\Lambda _{{\rm b}}/\Lambda _{{\rm f}}$ over 3 in
the O(4) case, we do not have solution of Eq.~(\ref{eq:GE1-exp}) 
with $F_\pi $
fitted to the experimental value. Calculations in the baryon sector of the
\begin{figure}[tbp]
\xslide{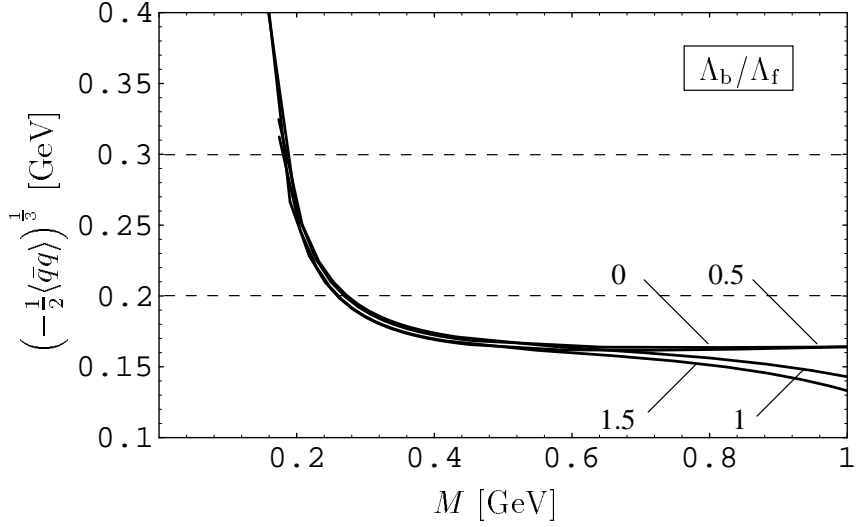}{7.5cm}{80}{280}{500}{490}
\caption{The quark condensate $\left(-{\frac{1}{2}}\langle \bar q q
\rangle\right)^{\frac{1}{3}}$ with proper-time quark-loop regularization as
a function of $M$  for different values of $%
\Lambda_{{\rm b}} / \Lambda_{{\rm f}} $. The curves correspond to 
 ${\Lambda_{{\rm b}}/ \Lambda_{{\rm f}} } = $ 0, 0.5, 1, 1.5 . The dashed
lines mark the empirical bounds. }
\label{fig:MQQPT}
\end{figure}
Nambu--Jona-Lasinio model with one quark and zero boson loops~\cite
{Meissner91,Christo:ff} show that best results for baryonic properties are
obtained for $M$ in the range 400--500~MeV, {\em i.e.}\ for values of the
quark cut-off near the critical value. Although this picture may change when
meson-loop effects are included, we note that the value $M$ 
for the critical value of the quark cut-off changes very little.
For $\Lambda _{{\rm b}}\sim \Lambda _{{\rm f}}$ the fermion cut-off
increases by about 30\%.
\begin{figure}[tbp]
\xslide{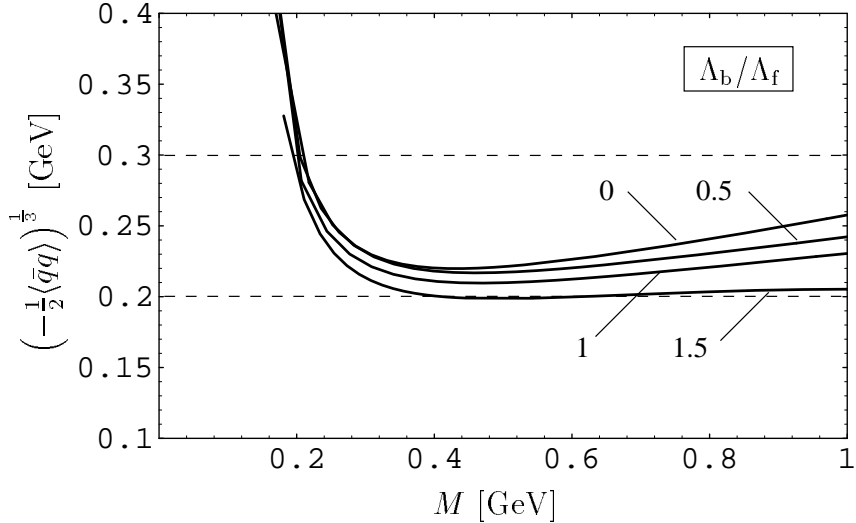}{7.5cm}{80}{280}{500}{490}
\caption{The quark condensate $\left(-{\frac{1}{2}}\langle \bar q q
\rangle\right)^{\frac{1}{3}}$ with O(4) quark-loop regularization as a
function of $M$  for different values of $%
\Lambda_{{\rm b}} / \Lambda_{{\rm f}} $. The curves correspond to 
 ${\Lambda_{{\rm b}}/ \Lambda_{{\rm f}} } = $ 0, 0.5, 1.5, 1 . The dashed
lines mark the empirical bounds.}
\label{fig:MQQO4}
\end{figure}

Comparing the results of the two fermion-loop regularization procedures we
can see that although the critical values of $\Lambda_{{\rm f}}$, below
which there is no spontaneous chiral symmetry breaking are quite different,
the general behavior of $M$ as a function of the fermionic and mesonic
cut-offs is very similar.
\begin{figure}[tbp]
\xslide{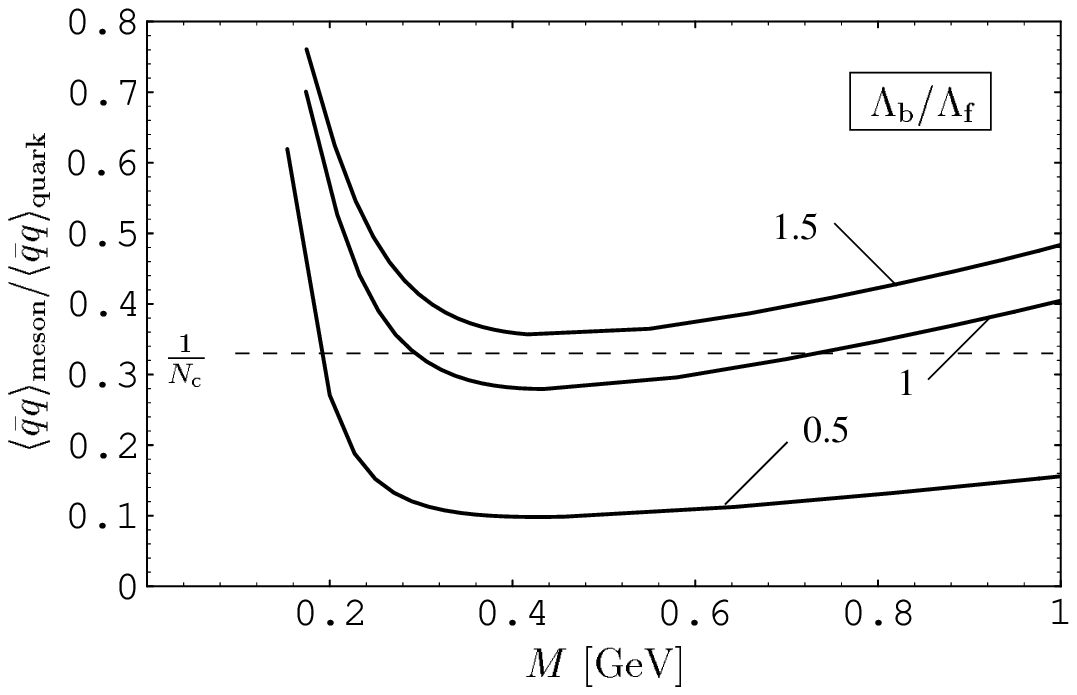}{7.5cm}{80}{280}{500}{490}
\caption{The ratio of the meson-loop to the
quark-loop contribution to  $\langle \bar qq\rangle$ as a function of $M$
obtained with proper-time fermion-loop regularization. The curves
correspond to ${\Lambda_{{\rm b}}/ \Lambda_{{\rm f}} }
= $0.5, 1, 1.5 .}
\label{fig:MQQrPT}
\end{figure}

The quark condensate is given in terms of $M$ and the
quark-quark coupling constant by Eq.~(\ref{eq:qq1}). The results in the
chiral limit are shown in Figs.~\ref{fig:MQQPT} and~\ref{fig:MQQO4} for the
proper-time and O(4) fermion-loop regularizations. They are plotted as a
function of $M$ for the four different ratios $%
\Lambda _{{\rm b}}/\Lambda _{{\rm f}}$. The empirical value for the quark
condensate is obtained in chiral perturbation theory using the value of the
quark current mass (Gell-Mann--Oakes--Renner relation)~\cite{Gasser82} or
from QCD sum-rules. 
The phenomenological uncertainty~(\ref{eq:qq-exp}) is, however, quite wide.
With both the proper-time and O(4) cut-offs there is a plateau for $M$
between 0.3 and 0.6~GeV and the corresponding value of $\langle \bar qq
\rangle$ depends little on $\Lambda_{{\rm b}} / \Lambda_{{\rm f}}$. The
value in the proper-time case (Fig.~\ref{fig:MQQPT}) is underestimated. This
problem is known from the one-quark-loop calculations, where better
agreement can be obtained using a generalized two-parameter proper-time
regularization function. In the case of the O(4) quark-loop regularization
(Fig.~\ref{fig:MQQO4}) our results for the quark condensate are in the
phenomenological bounds for values of $M$ in the plateau region. This shows
that the quark condensate is sensitive to the particular quark-loop
regularization procedure, a feature well known from the calculations in the
one-quark-loop approximation.
\begin{figure}[tbp]
\xslide{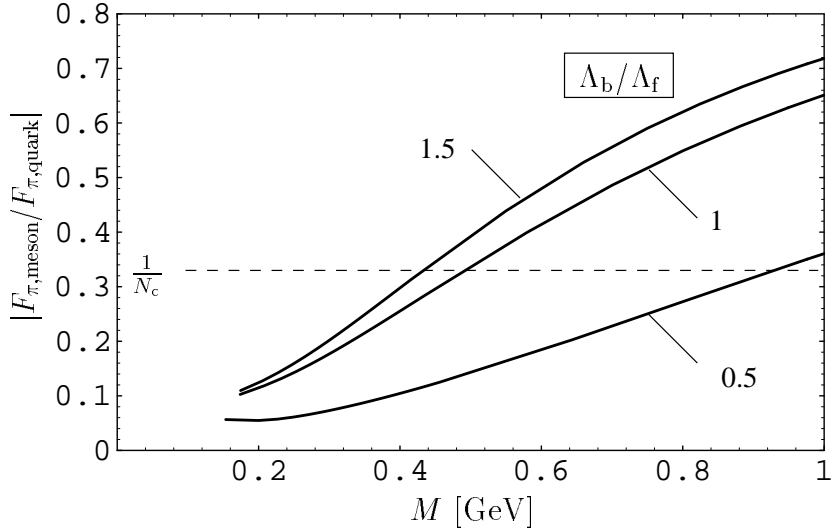}{7.5cm}{80}{280}{500}{490}
\caption{The ratio of the meson-loop to the
quark-loop contribution to  $F_\pi$ as a function of $M$
obtained with proper-time fermion-loop regularization. The curves
correspond to ${\Lambda_{{\rm b}}/ \Lambda_{{\rm f}} }
= $0.5, 1, 1.5 .}
\label{fig:MfpirPT}
\end{figure}

In Fig.~\ref{fig:MQQrPT} we plot the ratio of the meson-loop to the
quark-loop contribution to $\langle \bar qq\rangle$ as a function of $M$ for
the proper-time quark-loop regularization. The relative contribution of the
meson loop increases with increasing $\Lambda_{{\rm b}}/ \Lambda_{{\rm f}}$.
For values of the parameters in the range $0.3<M<0.6$~GeV and $\Lambda_{{\rm %
b}}/ \Lambda_{{\rm f}}<1$ we have $\langle \bar qq\rangle_{{\rm meson}} /
\langle \bar qq\rangle_{{\rm quark}}<40$\%, {\em i.e.} of the order of $%
\frac{1}{N_{{\rm c}}}$ as expected.

Looking at Figs.~\ref{fig:MQQPT} and~\ref{fig:MQQrPT} we see that as ${%
\Lambda_{{\rm b}}/ \Lambda_{{\rm f}} }$ increases the meson-loop
contribution to the quark condensate increases, while the quark-loop
contribution decreases. The sum of the contributions remains roughly
constant. Eq.~(\ref{eq:qq1}) shows that the inclusion of the meson-loops
does not significantly change the four-quark coupling constant ${\frac{1}{%
a^2}}$.

Let us now come back to the $N_{{\rm c}}$-counting scheme, which we
extensively used throughout the meson-loop calculations presented above. We
have calculated the meson-loop corrections to the quark condensate,
the pion mass and the pion decay constant, showing that they are one
order in $N_{{\rm c}}$ suppressed. However, the expansion in powers of ${%
\frac 1{N_{{\rm c}}}}$ should be used with care since $N_{{\rm c}}=3$.
Furthermore, the meson-loop corrections depend on the mesonic cut-off $%
\Lambda _{{\rm b}}$ and are infinite for $\Lambda _{{\rm b}}\to \infty $.
Hence it is important to check how well does the $N_{{\rm c}}$-counting
scheme work for the particular values of the parameters.

In the previous section we have compared the meson-loop and quark-loop
contributions to $\langle \overline q q \rangle$ and found that for
reasonable values of the parameters the ratio $\langle \overline q q
\rangle_{{\rm meson}} / \langle \overline q q \rangle_{{\rm quark}}$ is in
the expected range. Here we compare the meson- and quark-loop contributions
to $F_\pi$. Since we have fixed the parameters in order to reproduce the
experimental value of the pion decay constant the sum of both contributions
gives $F_{\pi,{\rm quark}} + F_{\pi,{\rm meson}} = 93$~MeV, but their
relative contributions depend on the choice of parameters. The meson-loop
contribution is negative. In Fig.~\ref{fig:MfpirPT} we show the ratio $%
|F_{\pi,{\rm meson}} / F_{\pi,{\rm quark}}|$ with proper-time quark-loop
regularization. The relative contribution of the meson-loop grows with
increasing $M$. It is larger for larger values of $%
\Lambda_{{\rm b}}/\Lambda_{{\rm f}}$. As we mentioned in the previous
section, the empirical value of $F_\pi$ can be only reproduced for $\Lambda_{%
{\rm b}}/\Lambda_{{\rm f}}$ less than about 2, which sets an upper bound for
the meson-loop cut-off. For values of $\Lambda_{{\rm b}}$ in this range and $%
M<0.6$~GeV the meson-loop correction does not exceed 40\% of the $N_{{\rm c}%
} $-leading quark-loop term. This is important because it keeps the $\frac{1 
}{N_{{\rm c}}}$ corrections in the expected range. It justifies {\em a
posteriori} the use of the $N_{{\rm c}}$-counting scheme. The above
observations are also true for the quark condensate (see Fig.~\ref
{fig:MQQrPT}) as well as for the O(4) fermionic regularization. In the case
when $\Lambda _{{\rm b}}\simeq \Lambda _{{\rm f}}$ and $0.3 < M < 0.6$~MeV
we find that the meson-loop contributions are of the order of ${\frac 1{N_{%
{\rm c}}}}={\frac 13}$ compared to the leading-order contributions.

\section{Conclusion}

\label{sec:concl} We have shown that the effective action formalism in the
Nambu--Jona-Lasinio model leads to a symmetry-conserving approximation,
allowing us include consistently meson-loop effects while preserving the
usual properties associated with spontaneous chiral symmetry breaking (the
Goldstone theorem, the Gell-Mann--Oaks--Renner relation, the form
of the leading nonanalytic terms in the chiral expansion of various
quantities).
Indeed, meson-loop effects will destroy these properties unless both the gap
equation and meson propagators are treated consistently. This has already
been noticed in Ref.~\cite{Tegen95}, where Feynman diagrams were used. In
our approach the conservation of the symmetry properties is a natural
consequence of the consistent application of the $\frac{1}{N_{{\rm c}}}$
expansion. We have found that though meson-loop contributions are 
$\frac{1}{N_{{\rm c}}}$-suppressed as compared to the leading-order quark-loop
contributions they lead to substantial corrections. For the physically
reasonable values of the model 
parameters the effects of meson loops to the pion
decay constant and to the quark condensate are of the expected order of~30\%.

\section*{Acknowledgment}

The authors acknowledge the support of the Volkswagen Foundation (EN, CC),
the Bulgarian National Science Foundation, contract $\Phi$-406 (EN, CC), of
the Polish State Committee of Scientific Research, grant 2~P03B~188~09 (WB),
and of the Alexander von Humboldt Foundation (GR, WB), as well as partial
support by COSY, BMBF, PROCOPE, and 
Stiftung f\"ur Deutsch-Polnische Zusammenarbeit.

\appendix

\section{Quark-loop meson vertices}

\label{app:Seff-exp} In our meson-loop calculations we need the meson
vertices which couple different number of mesons (we need up to four)
through a quark loop. They are given by the variations of the effective
action with respect to the meson fields. To obtain these we first expand the
action in powers of the meson field fluctuations around the stationary
point. We perform the expansion for the proper-time regularized effective
action. Using the proper-time expressions we can easily obtain the
four-momentum regularized expressions as well.

We start from the proper-time regularized effective action 
\begin{equation}  \label{eq:SeffPTf}
I = {\frac{N_{{\rm c}} }{2}} {\rm Tr} \int_{\Lambda_{{\rm f}}^{-2}}^\infty {%
\frac{ds }{s}}\, {\rm e}^{-s D^\dagger D}+ {\frac{a^2 }{2}}\, \Phi^2 - a^2
m\,\Phi_0,
\end{equation}
where $D^\dagger D$ is given by 
\begin{equation} 
D^{\dagger }D=\partial _\mu \partial _\mu +i\gamma _\mu \Gamma _\alpha\left(
\partial _\mu \Phi _\alpha\right)
+ \Phi^2\,. 
\label{DDf1}
\end{equation}
The fields $\Phi$ are now fluctuating around the stationary-point constant
fields which are of the form $\Phi^{{\rm st}}=\{S,0,0,0\}$ 
\begin{equation}
\Phi = \Phi^{{\rm st}} + \delta\Phi\, .
\end{equation}
Expanding $D^\dagger D$ in terms of the meson field fluctuations we obtain 
\begin{equation}
D^\dagger D \equiv G_0^{-1} + V\, ,
\end{equation}
where 
\begin{equation}
G_0^{-1} = \partial^\mu \partial_\mu + {\Phi^{{\rm st}}}^2
\end{equation}
is the zero-order term in $\delta\Phi$ which is diagonal in momentum space
and the fluctuation term is given by 
\begin{eqnarray}
V &=&V^{(1)} + V^{(2)} \, ,  \nonumber \\
V^{(1)} &=& 2 \Phi^{{\rm st}}_a \delta\Phi_a + 
{\rm i}\gamma _\mu \Gamma _\alpha\left(\partial _\mu \delta\Phi _\alpha\right)
\,,  \nonumber \\
V^{(2)} &=& (\delta\Phi)^2  \label{eq:Vexp}
\end{eqnarray}
and consists of terms of first, $V^{(1)}$, and second order, $V^{(2)}$, in
the field fluctuations. Next we expand the exponent in the fermionic part of
the effective action (\ref{eq:SeffPTf}) in powers of $V$ using the
Feynman-Schwinger-Dyson formula and the cyclic property of the trace 
\begin{eqnarray}
&&I_{{\rm f}} = {\frac{N_{{\rm c}} }{2}}{\rm Tr} \int_{\Lambda_{{\rm f}%
}^{-2}}^\infty {\frac{ds }{s}} \left({\rm e}^{-s D^\dagger D}\right) 
\nonumber \\
&&={\frac{N_{{\rm c}} }{2}}{\rm Tr} \int_{\Lambda_{{\rm f}}^{-2}}^\infty {%
\frac{ds }{s}}\left\{ {\rm e}^{-s G_0^{-1}} - s V {\rm e}^{-s G_0^{-1}} + {%
\frac{s^2 }{2}} \int_0^1 du\, V {\rm e}^{-s (1-u) G_0^{-1}} V {\rm e}^{-s u
G_0^{-1}}\right.  \nonumber \\
&&- {\frac{s^3 }{3}} \int_0^1 du \int_0^{1-u} dv\, V {\rm e}^{-s (1-u-v)
G_0^{-1}} V {\rm e}^{-s u G_0^{-1}} V {\rm e}^{-s v G_0^{-1}}
\label{eq:SeffFVV} \\
&&\left. + {\frac{s^4 }{4}} \int_0^1 du \int_0^{1-u} dv\! \int_0^{1-u-v}\!
dw\, V {\rm e}^{-s (1-u-v-w) G_0^{-1}} V {\rm e}^{-s u G_0^{-1}} V {\rm e}%
^{-s v G_0^{-1}} V {\rm e}^{-s w G_0^{-1}} \right\}  \nonumber
\end{eqnarray}
The first term in this expansion (zero order in $V$) gives the quark
contribution to the one-quark-loop effective action (\ref{eq:SeffPT}). The
term linear in $\delta\Phi$ cancels with the mesonic terms in the expansion
of the action since we expand around the stationary point.

The term quadratic in $\delta\Phi$ acquires contributions from the linear
and quadratic in $V$ terms as well as from the mesonic part of the effective
action. Evaluating it in momentum space and taking the second variation with
respect to the meson fields we obtain 
\begin{eqnarray}
S_{ab} &\equiv& {\frac{\delta^2 I }{\delta\Phi_\alpha(q_1)
\delta\Phi_\beta(q_2)}}  \nonumber \\
&=& \delta_{\alpha\beta} \delta (q_1+q_2) \left\{ 4 N_{{\rm c}} \left[
f(S,q_1) (q_1^2 + 4 S^2 \delta_{\alpha 0}) -2 g(S) \right] +a^2\right\}\, ,
\label{eq:Sab}
\end{eqnarray}
where the functions $f$ and $g$ are calculated in App.~\ref{app:freg}. 

The third order term in the expansion of the action $I$ in powers of the
field fluctuations $\delta\Phi$ acquires contributions from the second and
third order in $V$ terms in (\ref{eq:SeffFVV}) 
\begin{eqnarray}  \label{eq:S3B}
I^{(3)} &=& I^{(3A)} + I^{(3B)}\, , \\
I^{(3A)} &=& N_{{\rm c}} \sum_{C_V}^2 {\rm Tr} \int_{\Lambda_{{\rm f}%
}^{-2}}^\infty ds\, {\frac{s }{4}} \int_0^1 du\, V^{(1)} {\rm e}^{-s (1-u)
G_0^{-1}} V^{(2)} {\rm e}^{-s u G_0^{-1}}, \\
I^{(3B)} &=& -N_{{\rm c}} {\rm Tr} \int_{\Lambda_{{\rm f}}^{-2}}^\infty ds\, 
{\frac{s^2 }{6}} \int_0^1 du \int_0^{1-u} dv\, V^{(1)} {\rm e}^{-s (1-u-v)
G_0^{-1}}  \nonumber \\
&&\quad\times V^{(1)} {\rm e}^{-s u G_0^{-1}} V^{(1)} {\rm e}^{-s v
G_0^{-1}} ,  \nonumber \\
\end{eqnarray}
where $\displaystyle\sum_{C_V}$ denotes the sum over the permutations of $%
V^{(1)}$ and $V^{(2)}$. Varying this expressions with respect to the meson
fields we obtain the corresponding three-leg quark-meson vertex functions in
momentum space 
\begin{eqnarray}
S_{abc}^{(A)} &\equiv& {\frac{\delta^3 I^{(3A)} }{\delta\Phi_\alpha(q_1)
\delta\Phi_\beta(q_2) \delta\Phi_\gamma(q_3)}} = 8N_{{\rm c}} \sum_P^{3!}
\delta(q_1+q_2+q_3) \Phi^{{\rm st}}_\alpha \delta_{\beta\gamma}  \nonumber \\
&&\times \int_{\Lambda_{{\rm f}}^{-2}}^\infty ds\, s \int_0^1 du \int {\frac{%
d^4 k }{(2\pi)^4}}\, {\rm e}^{-s(1-u)((k+q_1)^2+S^2)} {\rm e}^{-su(k^2+S^2)},
\label{eq:SabcA}
\end{eqnarray}
\begin{eqnarray}
S_{abc}^{(B)} &\equiv& {\frac{\delta^3 I^{(3B)} }{\delta\Phi_\alpha(q_1)
\delta\Phi_\beta(q_2) \delta\Phi_\gamma(q_3)}} = - {\frac{8N_{{\rm c}}}{3}}
\sum_P^{3!} \delta(q_1+q_2+q_3)  \nonumber \\
&&\times \left( 4 \Phi^{{\rm st}}_\alpha \Phi^{{\rm st}}_\beta \Phi^{{\rm st}%
}_\gamma + \delta_{\alpha\beta}\Phi^{{\rm st}}_\gamma\, q_1\cdot q_{2} +
\delta_{\beta\gamma}\Phi^{{\rm st}}_\alpha\, q_2\cdot q_{3} +
\delta_{\alpha\gamma}\Phi^{{\rm st}}_\beta\, q_1\cdot q_{3}\right)  \nonumber
\\
&&\times \int_{\Lambda_{{\rm f}}^{-2}}^\infty ds\, s^2 \int_0^1\! du
\int_0^{1-u}\! dv \int {\frac{d^4 k }{(2\pi)^4}}\,  \nonumber \\
&&\times {\rm e}^{-s(1-u-v)((k+q_1)^2+S^2)} {\rm e}^{-su((k+q_1+q_2)^2+S^2)} 
{\rm e}^{-sv(k^2+S^2)} ,  \label{eq:SabcB}
\end{eqnarray}
where we have introduced the notation $\displaystyle\sum_P$ for the sum over
the permutations of the couples of meson-field indices and momenta $%
\{(\alpha,q_1),(\beta,q_2),(\gamma,q_3),\dots\}$. Summing the two
contributions we obtain finally 
\begin{equation}
S_{abc} = S_{abc}^{(A)} + S_{abc}^{(B)} \;. \label{eq:SabcS}
\end{equation}

The fourth order term in the expansion of ${I}$ in powers of the field
fluctuations $\delta\Phi$ acquires contributions from the second, third and
fourth order in $V$ terms in (\ref{eq:SeffFVV}) 
\begin{eqnarray}  \label{eq:S4A}
I^{(4)} &=& I^{(4A)} + I^{(4B)} + I^{(4C)}\, , \\
I^{(4A)} &=& N_{{\rm c}} {\rm Tr} \int_{\Lambda_{{\rm f}}^{-2}}^\infty ds\, {%
\frac{s }{4}} \int_0^1 du\, V^{(2)} {\rm e}^{-s (1-u) G_0^{-1}} V^{(2)} {\rm %
e}^{-s u G_0^{-1}}, \\
I^{(4B)} &=& - N_{{\rm c}} \sum_{C_V}^3 {\rm Tr} \int_{\Lambda_{{\rm f}%
}^{-2}}^\infty ds\, {\frac{s^2 }{6}} \int_0^1 du \int_0^{1-u} dv\,  \nonumber
\\
&&\times V^{(2)} {\rm e}^{-s (1-u-v) G_0^{-1}} V^{(1)} {\rm e}^{-s u
G_0^{-1}} V^{(1)} {\rm e}^{-s v G_0^{-1}},  \label{eq:S4B} \\
I^{(4C)} &=& N_{{\rm c}} {\rm Tr} \int_{\Lambda_{{\rm f}}^{-2}}^\infty ds\, {%
\frac{s^3 }{8}} \int_0^1 du \int_0^{1-u} dv \int_0^{1-u-v} dw\,  \nonumber \\
&&\times V^{(1)} {\rm e}^{-s (1-u-v-w) G_0^{-1}} V^{(1)} {\rm e}^{-s u
G_0^{-1}} V^{(1)} {\rm e}^{-s v G_0^{-1}} V^{(1)} {\rm e}^{-s w G_0^{-1}}.
\label{eq:S4C}
\end{eqnarray}
We evaluate the traces in momentum space and vary with respect to the meson
fields in order to obtain the corresponding contributions to the four-leg
quark-meson vertex 
\begin{eqnarray}
&&S_{abcd}^{(A)}\equiv {\frac{\delta^4 I^{(4A)} }{\delta\Phi_\alpha(q_1)
\delta\Phi_\beta(q_2) \delta\Phi_\gamma(q_3) \delta\Phi_\delta(q_4)}} = 2 N_{%
{\rm c}} \sum_P^{4!} \delta(q_1+q_2+q_3+q_4)  \nonumber \\
&&\quad\times \delta_{\alpha\beta} \delta_{\gamma\delta} \int_{\Lambda_{{\rm %
f}}^{-2}}^\infty ds\,s \int_0^1 du \int {\frac{d^4 k }{(2\pi)^4}}\, {\rm e}%
^{-s(1-u)((k+q_1+q_2)^2+S^2)} {\rm e}^{-su(k^2+S^2)}\, ,  \label{eq:SabcdA}
\end{eqnarray}
\begin{eqnarray}
&&S_{abcd}^{(B)}\equiv {\frac{\delta^4 I^{(4B)} }{\delta\Phi_\alpha(q_1)
\delta\Phi_\beta(q_2) \delta\Phi_\gamma(q_3) \delta\Phi_\delta(q_4)}} = - 4
N_{{\rm c}} \sum_P^{4!} \delta(q_1+q_2+q_3+q_4)  \nonumber \\
&&\quad\times \delta_{\alpha\beta} \delta_{\gamma\delta} (q_3\cdot q_{4} + 4
S^2 \delta_{\gamma 0}) \int_{\Lambda_{{\rm f}}^{-2}}^\infty ds\,s^2 \int_0^1
du \int_0^{1-u} dv \int {\frac{d^4 k }{(2\pi)^4}}  \nonumber \\
&&\quad\times {\rm e}^{-s(1-u-v)((k+q_1+q_2)^2+S^2)} {\rm e}%
^{-su((k+q_1+q_2+q_3)^2+S^2)} {\rm e}^{-sv(k^2+S^2)}\, ,  \label{eq:SabcdB}
\end{eqnarray}
\begin{eqnarray}
&&S_{abcd}^{(C)}\equiv {\frac{\delta^4 I^{(4C)}_{{\rm eff}} }{%
\delta\Phi_\alpha(q_1) \delta\Phi_\beta(q_2) \delta\Phi_\gamma(q_3)
\delta\Phi_\delta(q_4)}} = N_{{\rm c}} \sum_P^{4!} \delta(q_1+q_2+q_3+q_4)\,
\nonumber \\
&&\quad\times \biggl[16 \Phi^{{\rm st}}_\alpha \Phi^{{\rm st}}_\beta \Phi^{%
{\rm st}}_\gamma \Phi^{{\rm st}}_\delta + 4 \sum_{C_{V^{\prime}}}^6
(\delta_{\mu\nu} \Phi^{{\rm st}}_\kappa\Phi^{{\rm st}}_\chi\, q_i\cdot q_j)
+ (\delta^{\alpha\beta}\delta^{\gamma\delta}\!\! -
\delta^{\alpha\gamma}\delta^{\beta\delta}\!\! +
\delta^{\alpha\delta}\delta^{\beta\gamma})\,  \nonumber \\
&&\quad \times \bigl((q_1\cdot q_2)(q_3\cdot q_4) - (q_1\cdot q_3)(q_2\cdot
q_4) + (q_1\cdot q_4)(q_2\cdot q_3)\bigr)\biggr]  \nonumber \\
&&\quad\times \int_{\Lambda_{{\rm f}}^{-2}}^\infty ds\,s^3 \int_0^1 du
\int_0^{1-u} dv \int_0^{1-u-v} dw \int {\frac{d^4 k }{(2\pi)^4}}\, {\rm e}%
^{-s(1-u-v-w)((k+q_1)^2+S^2)}  \nonumber \\
&&\quad\times {\rm e}^{-su((k+q_1+q_2)^2+S^2)} {\rm e}%
^{-sv((k+q_1+q_2+q_3)^2+S^2)} {\rm e}^{-sw(k^2+S^2)}\, ,  \label{eq:SabcdC}
\end{eqnarray}
where in the last contribution $\displaystyle\sum_{C_{V^{\prime}}}^6$ stays
for the sum over combinations, originating from different orderings of the
two terms in $V^{(1)}$~(\ref{eq:Vexp}) 
\begin{eqnarray}
&&\sum_{C_{V^{\prime}}}^6 (\delta_{\mu\nu} \Phi^{{\rm st}}_\kappa\Phi^{{\rm %
st}}_\chi\, q_i\cdot q_j) \equiv \delta_{\alpha\beta} \Phi^{{\rm st}%
}_\gamma\Phi^{{\rm st}}_\delta\, q_1\cdot q_{2} + \delta_{\alpha\gamma}
\Phi^{{\rm st}}_\beta\Phi^{{\rm st}}_\delta\, q_1\cdot q_{3}  \label{eq:CV6}
\\
&&\; + \delta_{\alpha\delta} \Phi^{{\rm st}}_\beta\Phi^{{\rm st}}_\gamma\,
q_1\cdot q_{4} + \delta_{\beta\gamma}\Phi^{{\rm st}}_\alpha\Phi^{{\rm st}%
}_\delta\, q_2\cdot q_{3} + \delta_{\beta\delta} \Phi^{{\rm st}}_\alpha\Phi^{%
{\rm st}}_\gamma\, q_2\cdot q_{4} + \delta_{\gamma\delta} \Phi^{{\rm st}%
}_\alpha\Phi^{{\rm st}}_\beta\, q_3\cdot q_{4}.  \nonumber
\end{eqnarray}
Summing up all contributions we obtain finally 
\begin{equation}
S_{abcd} = S_{abcd}^{(A)} + S_{abcd}^{(B)} + S_{abcd}^{(C)} \;.
\label{eq:SabcdS}
\end{equation}

\section{Regularization functions}

\label{app:freg} Here we calculate the regularization functions for the
quark-loop vertices in the diagrams for the gap equation and
pion propagator. We start by evaluating the regulators in the proper-time
scheme. The O(4) regulators can be then easily obtained using the
intermediate results of the proper-time case.

\subsection{Proper-time regularization}

\label{app:fregPT} We start by evaluating the function $g$ emerging in the
one-quark-loop gap equation. It is given by 
\begin{eqnarray}
g(S) &\equiv& \int {\frac{d^4 k }{(2\pi)^4}} \int_{\Lambda_{{\rm f}%
}^{-2}}^\infty ds\, {\rm e}^{-s \left(k^2 + S^2\right)}  \nonumber \\
&=& {\frac{1 }{16\pi^2}} \left(\Lambda_{{\rm f}}^2 {\rm e}^{-{\frac{S^2}{%
\Lambda_{{\rm f}}^2}}} - S^2 E_1\left({\frac{S^2}{\Lambda_{{\rm f}}^2}}%
\right)\right),  \label{eq:gMPT}
\end{eqnarray}
where $E_1$ is the exponential integral defined by 
\begin{equation}
E_n(x) \equiv \int_1^\infty dt \, {\frac{{\rm e}^{-xt}}{t^n}}\, .
\label{eq:expint}
\end{equation}
The function $f$ is defined by 
\begin{eqnarray}
f(S,q^2) &\equiv& \int {\frac{d^4 k }{(2\pi)^4}} \int_{\Lambda_{{\rm f}%
}^{-2}}^\infty ds\, s \int_0^1 du\, {\rm e}^{-s (1-u) \left((k+{\frac{q}{2}}%
)^2 + S^2\right)} {\rm e}^{-s u \left((k-{\frac{q}{2}})^2 + S^2\right)} 
\nonumber \\
&=& {\frac{1 }{16\pi^2}} \int_0^1 du\, E_1 \left( {\frac{S^2 + u(1-u)q^2 }{%
\Lambda_{{\rm f}}^2}} \right).  \label{eq:fMPT}
\end{eqnarray}

\subsection{O(4) regularization}

\label{app:fregO4} Here we define the quark-loop O(4) regularization. The
running four-mo\-men\-tum of the quark loop is limited in the following
manner. We obtain the quark-loop vertex functions in our O(4) regularization
scheme using the intermediate results from the proper-time calculation.
After having completed the square for the fermion-loop four-momentum $k$ in
the proper-time expressions we take the limit in the {\em proper-time}
fermionic cut-off $\Lambda_{{\rm f}}^{PT}\to \infty$ using the identity 
\begin{equation}
\lim_{\Lambda_{{\rm f}}^{PT}\to \infty} \int_{{(\Lambda_{{\rm f}}^{PT})}%
^{-2}}^\infty ds\, s^n {\rm e}^{-sA} = {\frac{n! }{A^{n+1}}}, \quad {\rm for}%
\ n\ge 0,\ A>0\, .
\end{equation}
Then we cut off the integral at some O(4) cut-off $\Lambda_{{\rm f}}^{O(4)}$
and obtain the O(4) regularization functions. Note that in the following
expressions both the $k^2$- and $u$-integrals can be taken analytically.
This leads, however, to quite lengthy sums of logarithms and rational
functions, and for the sake of clarity we present the regularization
functions keeping the integrals.

The function $g(S)$ can be easily obtained from the O(4)-regularized action 
\begin{equation}
g(S) = {\frac{1 }{16 \pi^2}} \int_0^{\Lambda_{{\rm f}}^2} d(k^2)\, {\frac{k^2%
}{k^2 + S^2}} ,  \label{eq:gMO4A}
\end{equation}
Using Eq.~(\ref{eq:fMPT}) we obtain 
\begin{equation}
f(S,q^2) = {\frac{1 }{16 \pi^2}} \int_0^{\Lambda_{{\rm f}}^2} d(k^2)
\int_0^1 du\, {\frac{k^2 }{(k^2 + S^2 + u(1-u)q^2)^2}} \, ,  \label{eq:fMO4}
\end{equation}


\end{document}